\documentclass[default]{sn-jnl}

\usepackage{amssymb}

\usepackage{amsmath}
\usepackage[section]{placeins}

\jyear{2023}%

\theoremstyle{thmstyleone}%
\usepackage{float}
\theoremstyle{thmstyletwo}%

\theoremstyle{thmstylethree}%

\begin{document}

\title[Exploring RNNs trained for simple tasks.]{Exploring weight initialization, diversity of solutions, and degradation in recurrent neural networks trained for temporal and decision-making tasks.}


 \author[1,2,3]{Cecilia Jarne}\email{cecilia.jarne@unq.edu.ar}
  \author[1,2]{Rodrigo Laje}\email{rlaje@unq.edu.ar}
 \affil[1]{Universidad Nacional de Quilmes - Departamento de Ciencia y Tecnolog\'ia}
 \affil[2]{CONICET, Buenos Aires, Argentina}
 \affil[3]{ Center for Functionally Integrative Neuroscience, Department of Clinical Medicine, Aarhus University, Denmark}








\abstract{Recurrent Neural Networks (RNNs) are frequently used to model aspects of brain function and structure. In this work, we trained small fully-connected RNNs to perform temporal and flow control tasks with time-varying stimuli. Our results show that different RNNs can solve the same task by converging to different underlying dynamics and also how the performance gracefully degrades as either network size is decreased, interval duration is increased, or connectivity damage is increased. {For the considered tasks, we explored how robust the network obtained after training can be according to task parameterization. In the process, we developed a framework that can be useful to parameterize other tasks of interest in computational neuroscience}. Our results are useful to quantify different aspects of the models, which are normally used as black boxes and need to be understood in order to model the biological response of cerebral cortex areas.}

\keywords{Recurrent Neural Networks, connectivity, degradation}



\maketitle
\section{Introduction:}

Recurrent Neural Networks (RNN) emerged a few decades ago (see for example the founding works of Hopfield \cite{Hopfield3088}, Elman \cite{ELMAN1990179} and Funahashi and Nakamura \cite{DBLP:journals/nn/FunahashiN93, DBLP:journals/nn/Funahashi89}) to model the mechanisms of a variety of systems, such as brain processes \cite{Gerstner60,10.1371/journal.pcbi.1004792, 10.1371/journal.pcbi.1005175, 10.3389/fncom.2013.00070, nature_com, nature_03, reminton_02}, stability and control \cite{DENG2013281, DINH201444, 7966138, PhysRevLett.118.258101}, and in Machine Learning \cite{DBLP:journals/ijon/GallicchioMP17, nature_02, libro_keras, deep-learning}. 

Within the Machine Learning (ML) framework, the main objective is to produce efficient network topologies and training methods to solve computational problems. {The aim when implementing these models in Computational Neurosciences is to describe the recurrent connectivity of the brain cortex to understand the mechanisms that underlie different processes.} 

This paper takes relatively new training algorithms from the Machine Learning community and applies them to analyze computational problems related to temporal tasks relevant to Neurosciences. {In particular, we want models that could be used to study the problem of how cortical subpopulation processes underlie various temporal tasks inspired by aspects of cognition. For example, The Flip Flop is a working memory task considered previously in other works in the field of Computational Neuroscience \cite{DBLP:journals/neco/SussilloB13, BARAK20171}, as well as the Finite-duration oscillation generator.} 
{On the other hand, several versions and parametrizations of decision-making tasks were also considered and constitute an extended paradigm \cite{WANG2008215,Orhan2019,SOHN2019934, Jarne_2021, 10.1371/journal.pcbi.1009271}.}

Recurrent neural networks are powerful tools since it has been proven that, given enough units, they can be trained to approximate any dynamical system \cite{con_01, con_02, Gallacher2000, Chow2000}. It has been studied that RNNs can display complex dynamics including attractors, limit cycles, and chaos \cite{SUSSILLO2014156, doi:10.1146/annurev-neuro-092619-094115}.

It is well established that RNNs constitute a versatile model in neuroscience research. They can be trained to process temporal information and perform different tasks such as flow control and many kinds of operations that roughly represent computation in different brain areas. A simple RNN model could perform tasks that are similar to stimuli selection, gain modulation, and temporal pattern generation in the cortex \cite{nature_letter_hahnloser}.

Trained networks serve as a source of mechanistic hypotheses and also as a testing ground for data analyses that could link neural activity and behaviour. RNNs are {also} a valuable platform for theoretical investigation, and some aspects of RNN models are used to describe a great variety of experimental results observed in different studies, for example, working memory, motor control, temporal processing, and decision making \cite{susillo, carnevale, 10.3389/fncom.2017.00112, nature_04}. For instance, it has recently been studied that recurrent circuits in the brain may play a role in object identification \cite{PMID:31036945}.

Another aspect considered is the computational principles that allow decisions and actions regarding flexibility in time. In a recent review \cite{review-motor}, a dynamical system perspective is used to study such flexibility and shows how it can be achieved through manipulations of inputs and initial conditions.

Recent experimental recordings of neurons in the cerebral cortex also show complex temporal dynamics \cite{SUSSILLO2014156, DBLP:journals/neco/SussilloB13, nature_01, nature_com}, where different mechanisms of information flow control could be present and coexist.
Temporal aspects of an RNN constrain the parameters, topologies, and different parts of the computation. Those aspects deserve to be studied and will improve current neuronal models. Given the complexity of these systems, it is not surprising that there are still so many fundamental gaps in the theory of RNNs \cite{BARAK20171}, such as how RNNs control the flow of information \cite{10.3389/fncom.2011.00001}.

A recurrent neural network can be trained to reproduce a task considering two paradigms: On the one hand, we can have an intuition about the behaviour of the system. Here the network represents an abstract variable that obeys an equation of a low-dimensional dynamical system, and the dynamics can be translated into the connectivity of the network. The underlying mechanism is modelled using a low-dimensional dynamic system that is then implemented in a high-dimensional RNN \cite{BARAK20171}.


{An alternative paradigm involves building a functional RNN, which is utilized in machine learning as well as in Computational Neuroscience}. In this case, one presents to the network the information relevant to the task to be fulfilled. This is done in terms of input-output data but without any direct prescription on how to fulfil it, {other than the training algorithm of choice}. If the mechanisms implemented by the network can be conceptualized, the network analysis can become a method for generating hypotheses for future experiments and data analysis \cite{SUSSILLO2014156}. The present work is based on this approach. 
 
{Typically, a cognitive task consists of elementary sensory, cognitive, and motor processes}. We examined a set of neuroscience-inspired tasks: time reproduction \cite{Jazayeri2010}, oscillatory response \cite{10.3389/fncom.2013.00070, DBLP:journals/neco/SussilloB13}, a Flip Flop \cite{DBLP:journals/neco/SussilloB13} and a set of decision making boolean-like tasks. These simple tasks have been used in Cognitive and Computational Neuroscience in the context of RNN studies \cite{10.1371/journal.pcbi.1004792,SUSSILLO2014156,Yang2019ER}, among others with the aim to understand how different brain areas process information.

In particular, we have taken a set of temporal and decision-making tasks, parameterized them and trained recurrent neural networks to reproduce them. We used in the process a new open-source framework that we developed that could be useful for others to create their other tasks of interest, based on the ones that we parameterized to develop their models (See \textbf{Supplementary Information} \ref{sup-a}).

In this framework, we numerically studied  the properties of a simple model representing a cortical subpopulation of neurons that, after training, can perform tasks that are relevant to processing information and flow control.
 
{After training the networks, we analysed the various configurations that emerged}. We carry out also a set of studies on how the performance of the trained network degrades to show the scope and limitations of the model. These studies are related to the performance degradation either as network size is decreased or the time duration is increased, and what happens when the connectivity of trained networks is damaged.

 {We decided to consider simple tasks (based on the temporal and the decision-making paradigms)  to focus on the effects of training on the networks, and also to study which are the multiple mechanisms that can arise without considering tasks with contextual cues.}
 
The rest of the paper is organized as follows. In Section \ref{mat-met}, we describe the network model, training method, the details of the code implementation and task parameterization. In section \ref{results}, we explain each task and describe how those {were} implemented. Then, we present the results of the different studies that we performed. Section \ref{discu} we discuss the results obtained and finally in Section \ref{conclu} we present the conclusions.

\section{Methods} \label{mat-met}
\subsection{Model} \label{model}

{Motivated by models of interconnected neurons of the firing rate type \cite{PhysRevLett.61.259, WILSON19721,susillo_2009,10.1371/journal.pcbi.1006309}, we set out to study the dynamics of the discrete system as it was previously considered, for example, in \cite{doi:10.1073/pnas.1921609117, doi:10.1073/pnas.0804451105, Ceni2020}. We considered the discrete RNN given by:}

\begin{center}
\begin{equation}
\mathbf{H}(t)=\phi(\mathbf{W^{Rec}}\mathbf{H}(t)+\mathbf{W^{in}}\mathbf{X}(t))),
\label{eq-03}
\end{equation}
\end{center}

{ $\phi$ is the hyperbolic tangent. Very early works have proved that it is possible to use discrete-time RRNs to uniformly approximate a discrete-time state-space trajectory which is produced by either a dynamical system or a continuous-time function to any degree of precision \cite{488134, DBLP:journals/nn/FunahashiN93}. RNNs are universal approximators of dynamical systems.}

{The model was implemented in Python using Keras and TensorFlow. That allows us to make use of all the algorithms and optimizations developed by that ML community.}

\subsection{On training methods} \label{training}

There are a great variety of algorithms to train recurrent neural networks. In a recent work from the ML field, a very detailed survey on RNNs with new advances for training algorithms and modern recurrent architectures was presented in \cite{DBLP:journals/corr/abs-1801-01078}. 

{The training methods} for neural networks can be unsupervised or supervised. In this work, {we focused} on applying a supervised method.

The studies where some form of gradient descent is applied stand out in the literature of supervised methods. An example is the Reservoir computing paradigm with liquid- or echo-state networks \cite{doi:10.1162/089976602760407955}, where the modifications of the network weights are made in the output layer weights $  \mathbf{W ^ { out}} $.

One of the most outstanding methods was the one developed by Sussillo and Abbott \cite{susillo_2009}. They have developed a method called FORCE that allows them to reproduce complex output patterns, including human motion-captured data \cite{susillo_2009}. Modifications to the algorithm have also been applied successfully in various applications \cite{laje, 10.1371/journal.pone.0191527}.

A frequently used alternative is gradient descent with backpropagation for its calculation and then some optimization method for minimizing it. Given the recent advances in the implementation of this method with Open Source libraries, this is the method we explored.

{For the choice of the training algorithm we were inspired by \cite{DBLP:journals/corr/abs-1801-01078, TRISCHLER201667,russo-2018}. In \cite{TRISCHLER201667}, authors use the Adam method to train networks and perform numerical simulations. Adam is an algorithm for first-order stochastic gradient-based optimization of objective functions \cite{DBLP:journals/corr/KingmaB14}}. 

\subsection{Network implementation and training protocol}\label{protocol}

{We used a simple RNN model which is composed of three layers. One is the input, the second is the recurrent hidden layer, and the last is the output layer. The input layer has $k$ units, all of which are connected to all recurrent units ($N\times k$ input connections).} The inputs to this layer are a sequence of vectors through time $t$, whose components are $x_j(t) $, {with $j=1,\ldots k$}. Every input neuron from {the input layer} is connected to every neuron in the hidden layer. The connectivity weight between input neuron $i$ and hidden neuron $j$ is $ w_{ij}^{in}$ (matrix notation $ \mathbf{W^{in}}$). The hidden layer has $N$ recurrently connected units with the activity given by $h(t) = (h_1(t), h_2(t), ..., h_N(t)),$ with recurrent connectivity weights $ w_{ij}^{Rec}$  {(in matrix notation $ \mathbf{W^{Rec}}$)}. The initialization of the connections $ w_{ij}^{Rec}$ of hidden units is done with small non-zero elements to improve the overall performance of the training.  {The connectivity from the recurrent to the output layer} is represented by the $ \mathbf{W^{out}}$ matrix. {The connectivity from the recurrent to the output layer with weights $ w_{ij}^{out}$.}

In the present work, we implemented a recurrent neural network with $N=50$ hidden units, unless indicated. We used as activation function the hyperbolic tangent. 

The weight matrix $ \mathbf{W^{in}}$ was initialised randomly from the uniform distribution with the Glorot uniform initializer from Keras \cite{chollet2015keras}, which is the default selection. It draws samples from a uniform distribution within $[-limit, limit]$, where $limit = \sqrt{6 / (fan_{in} + fan_{out)}}$, ($fan_{in}$ is the number of input units and $fan_{out}$ is the number of output units \cite{tensorflow2015-whitepaper}). Recurrent weights matrix $ \mathbf{W^{Rec}}$ is initialized in two different ways during the study: either as a random matrix or as an orthogonal random matrix. The random matrix has elements drawn from a normal distribution with  standard deviation $\sigma= \frac{1}{\sqrt{N}}$ and mean $\mu = 0$, where N is the number of recurrent units. \cite{DBLP:journals/corr/SaxeMG13}. The orthogonal random matrices are created with Orthogonal initializer from Keras, where an orthogonal matrix is obtained by QR decomposition of a matrix of random numbers drawn from a normal distribution.

This selection allows us to implement Adam as a training method with low computational cost and diminish the impact of the vanishing gradient problem for this simple implementation. {The tasks can be learned in reasonable computational time and with good accuracy in 20 Epochs.}

\begin{figure}[htb!]
\begin{center}
\hspace{1cm}
\includegraphics[totalheight=9cm]{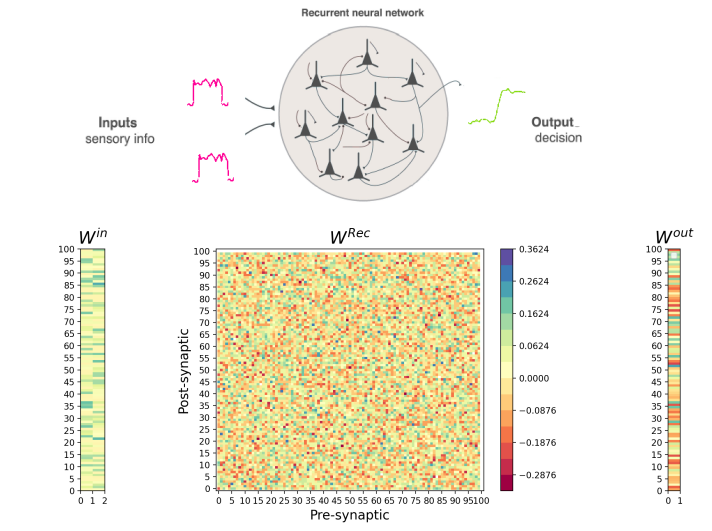}
\caption{Upper panel: neural network schema of our model with two units at the input to process the time series with arbitrary value and one output. Lower panel: example for a Neural Network connectivity matrices $\mathbf{ W^{in}}$, $\mathbf{ W^{Rec}}$, and $\mathbf{ W ^{out} }$ for a 100 unit recurrent network (arbitrary size) with two units at the input and one at the output. Rows represent the output connection (post-synaptic) of each unit or column (pre-synaptic).}
\label{fig_01}
\end{center}
\end{figure}

The upper panel of Figure \ref{fig_01} shows a schematic for the neural network model studied in this work. The input signal is a two-component vector with arbitrary time evolution. The output is the readout unit that provides the output decision. The lower panel shows and example for a Neural Network connectivity matrices $\mathbf{ W^{in}}$, $\mathbf{ W^{Rec}}$, and $\mathbf{ W ^{out} }$.

For instance, let us consider a network with 50 units that are further used in the analysis presented in Section \ref{results}. The units in the hidden layer are fully connected to each other, in the sense that the connectivity matrix $\mathbf{ W^{Rec}}$ is not sparse. Let us consider a two-input task, meaning that the input layer has two input units. The input layer $ \mathbf{W^{in}}$ will have $50\times 2$ connections to process a $2\times [length]$ vectors at the input, in general, we considered time series of 200 ms. The $\mathbf{ W^{Rec}}$ matrix has $50\times50$ connections that produce the temporal response of the 50 units with a vector of $50\times [200ms]$. At the output we have one unit representing the output decision that combines the 50 activity responses of the hidden units, using the $\mathbf{ W ^{out} }$ matrix of $1\times50$. It produces a vector $ \mathbf{Z(t)}$ of $1\times [200 ms]$.

The loss function used to train the model is the mean square error between the target function and the output of the network. It is defined as:

\begin{center}
\begin{equation}
{E(w)=\frac{1}{2}\sum_{t=1}\sum_{j=1} (\mathbf{Z_j}(t)-\mathbf{Z^{target} _j}(t))^2},
\label{eq-05b}
\end{equation}
\end{center}

where $Z^{target}_j(t)$ is the desired target function and $Z_j(t)$ is the actual output.

We trained the model with no less than 15000 input signal samples for all tasks described in the following section. We generated each sample of the training set as a noisy time series that may contain a square pulse with noise or only noise. Noise is an additive random variable drawn from a Gaussian distribution with zero mean, and its maximum  amplitude is 10\% the height of the pulse.

We also generated an additional set of input time series that we use to test after training. We check if the output response against the testing time series corresponds to the task for which the network has been trained.

In each experiment, we saved the initial pre-training configuration and the final instance of the network weights to study how the weight matrix changes during training. The aim of the training is to adjust all the parameters and obtain a network that can reproduce the task for which it was trained.

Framework and Code to train the networks and produce the Figures in this paper are open and available in the following repository:\\

 \url{https://github.com/katejarne/RNN\_study\_with\_keras}

\subsection{{General aspects of the tasks}} \label{tasks}

We considered that every input signal may have a square pulse of fixed duration, and the network responds according to the rule that it was trained for a given data set of the training sample.

{The stimulus widths and response times of the tasks are on the order of tens of milliseconds, part of the range of interest of signals processed by the cerebral cortex \cite{Goel20120460}. We considered the training data set as the set of input signals and target outputs with a low edge-triggered response to the input signals with a delay time of 20 ms. In each task presented in the following section} {(Figures from \ref{fig_02} to \ref{fig_08})}, we show a trained network responding to different testing samples. Each stimulus at the input is presented in green (input signal 1) and pink line (input signal 2) for those tasks with two inputs. The target is in a grey solid line, and the network response is in red. We considered a time series of between 200 ms and 500 ms length, but the length and the position of the stimulus are arbitrary.

\subsection{Description of the tasks considered in the study} \label{description}

The networks learn to decide which state the output signal should take in the face of the different stimuli presented at the input. In this sense, one task is defined as the different possible ways in which it can decide the output state based on input stimuli and the respective regimes that the output should take.

The motivation for the selection of the considered tasks is to simulate flow control processes that can occur in the cortex when receiving stimuli from other cortical or subcortical areas. In \cite{10.3389/fncom.2011.00001} the notion of gating was discussed as a mechanism capable of controlling the flow of information from one set of neurons to another. In the present work, the gating mechanisms are modelled using networks with a relatively small set of units.

It has been proposed that some sets of neurons in the brain could roughly function as gates \cite{10.3389/fncom.2011.00001}. The dynamics of trained networks for the Flip Flop task is also interesting, which is generally related to the concept of working memory. It has been previously studied in \cite{DBLP:journals/neco/SussilloB13, SUSSILLO2014156}, but in this case, with a more complex task referring to a 3-bit register called in the paper a 3-bit Flip Flop.

We focus on the study of networks trained for the following list of tasks related to the processing of stimuli as temporal inputs:

\begin{enumerate}
\item Time reproduction.
\item Basic logic gate operation: AND, OR, NOT, XOR.
\item Flip-Flop (1-bit memory storage).
\item Finite-duration oscillation.
\end{enumerate}

It should be noted that the tasks described in item 2 are not related to those made by static feedforward networks like \cite{libro_static} but to a network solving logic gates with time-varying inputs. We want to point out that, in item 3, we are not referring to the concept of ``Flip Flop neurons" such as the one proposed in \cite{7727548} but to a network learning the ``Flip-Flop rule" as in \cite{DBLP:journals/neco/SussilloB13} with two inputs. The focus is on the process of temporal signals similar to the XOR temporal task implemented in \cite{ELMAN1990179}. {In every task, the focus is on the processing of pulses (i.e., non-stationary signals)}. 

The RNN model emulates a ``cognitive-type" cortical circuit such as the prefrontal cortex, which receives converging inputs from multiple sensory pathways and projects downstream to other areas.
The chosen network architecture and size showed to be sufficient to learn all the tasks mentioned above.
We used dimensional reduction methods to study the inner state of the network during and after training and discuss, specifically, the results and observations regarding each task \cite{dim_red_nature}. In particular, PCA is the method that we chose because it has been widely used in the study of simulations, as well as experimental high dimensional neural space states \cite{10.1371/journal.pcbi.1002057}.

For the network implementation and training, we use the Keras libraries from \cite{chollet2015keras} and TensorFlow from \cite{tensorflow2015-whitepaper} as frameworks, instead of more traditional choices like Matlab from \cite{thompson1995image} or Theano from \cite{2016arXiv160502688short}, implemented in some works such as \cite{10.3389/fncom.2018.00083}. The reason for our selection is that these new scientific libraries are open-source, and their use is rapidly growing. In the case of Keras, it is the first time that it is used for such kind of study. In the case of TensorFlow, there are a few recent works that use it (for instance, see Ref \cite{williams}). 

\section{Results} \label{results}

The results of the numerical studies are shown in this section. First, we present in Section \ref{tasks-1} a description of every task and the parameterization used in the framework that we developed. This framework could be used to modify any of the seven tasks that we considered, but also it could it is possible to use it for developing different tasks of interest. In Section \ref{net-init} we discuss how different initialization schemas could improve the network training.

Section \ref{dyn} shows different aspects that we observed of the dynamics of the trained networks. Finally, Sections \ref{pulse_studies} and \ref{dam} show the result of different studies performed on the memory capacity, scale and damage of the trained networks.

To follow the examples shown in Section \ref{net-init} and Section \ref{dyn} (Figures \ref{fig_08_A} to \ref{fig_10_b}), the trained networks have been labelled. The labels include a number to identify the simulation number of the corresponding task and initial condition. These also allow identifying the data in the Supplementary Information to view the examples and the repository on Github.

{We have trained 20 networks for each task and initial condition. The trained networks shown in the paper and the Supplementary Information are available in .hdf5 files in the repository. The code provided also allows training more networks in each of the tasks described and also to plot network response to the testing stimuli.}

{For each of the cases presented, the networks exhibit the same response for the output, according to the training rule, each time the corresponding input is activated. We provide examples of the different stimuli combinations}. In addition to that, some of the tasks need two inputs (thus four different combinations to fully show its behaviour: AND, etc) while others only one input (thus only two different combinations: NOT for example), and others four combinations with history (Flip-Flop, not including repetitions of the same input). 

\subsection{Tasks}\label{tasks-1}

\subsubsection{Time reproduction} \label{memorizing}

Let's begin describing a simple temporal task considered in Section \ref{description}. In this task, when the network has a stimulus at the input (a Gaussian pulse with noise), it has to respond with a pulse at the output matching the input signal, and no response otherwise. 

We trained the network to produce an output pulse at a fixed time delay after an input pulse occurs, and output close to zero if there is no input pulse. Time delay is an arbitrary value parameter.

{This task is parameterized in this simple way to study how the network is capable of learning to respond after a certain interval of time. It is a simple task, much simpler than a standard perceptual decision-making task \cite{Britten4745}. In that case, the network must also learn to integrate a signal for a certain period of time and then a second time interval to respond. Here we are just training the network to learn to estimate one given time interval. We are interested in studying that process. In Section \ref{pulse_studies}, we show a study of how this temporal task is affected by varying the temporal interval or the size of the network. This is useful because it gives a minimal benchmark for time and network size for network hyperparameters studies and for other more complex tasks.}

\begin{figure}[htb!]
\begin{center}
\includegraphics[totalheight=3.5cm]{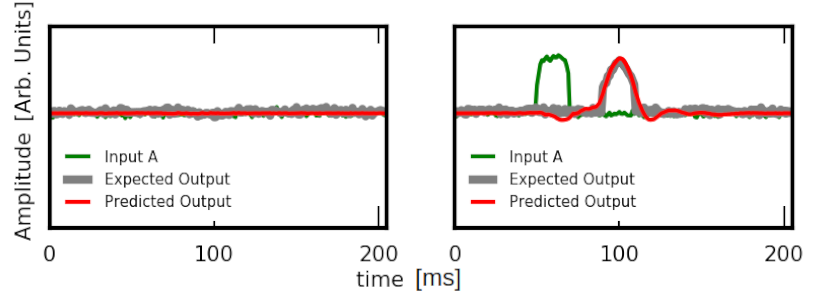}
\caption{{A trained neural network response to the testing input samples for the ``time reproduction'' task. The green line in the plot represents the input signal. The grey thick line is the target signal or expected output, and the red line is the network predicted output. Time is in ms and amplitude is in arbitrary units. }}
\label{fig_02}
\end{center}
\end{figure}

In Figure \ref{fig_02}, we show the simulations for two different input samples. The same neural network is considered and trained to memorize and reproduce a Gaussian pulse with noise at the input, after a 20 ms delay. If there is no pulse, the network must output a zero signal. This task was used in the scaling studies presented in Section \ref{pulse_studies}.

\subsubsection{Binary basic operations between input stimuli with AND, OR, XOR}\label{DM}

{These tasks presented here are a class of decision-making tasks. Here,} the network has to perform different binary-inspired operations with temporal stimuli at the input (or inputs). {As a result, the output will switch to a High value or keep a Low value} depending on the task and the presented input(s). The input stimuli are square signals with a duration of 20 ms and Gaussian noise of 10\% of the total amplitude. We considered our input data set random time series of 200 ms length with or without a pulse. For two-input tasks, the corresponding input pulses (if both are present) are simultaneous in time. The network has to decide the state of the output, which {should match} the training set rule for each considered task. We used 50-unit networks, and all networks were able to successfully reproduce all tasks after training.

For each of the tasks, we created the target output time series according to table \ref{tabla_and}. The truth tables for all logic operations are displayed in Table \ref{tabla_and}.

\begin{table}[h!]
\centering
\begin{tabular}{|c|c|c|c|c|c|}
\hline
\textbf{Input 1} & \textbf{Input 2} & \textbf{AND Output}  & \textbf{OR Output} &\textbf{XOR Output} \\ \hline
0                & 0                & 0     & 0 & 0        \\ \hline
0                & 1                & 0     & 1 & 1        \\ \hline
1                & 0                & 0     & 1 & 1         \\ \hline
1                & 1                & 1     & 1 & 0         \\ \hline
\end{tabular}

\label{tabla_and}
\vspace{0.2cm}
\caption{AND, OR and XOR states of the output with respect to the inputs states.}
\end{table}

{From inspecting the different realizations obtained, some general observations of these systems emerge when tasks are compared that are discussed in Section \ref{dyn}.}

\subsubsection{NOT task}\label{not}

The boolean NOT task consists of turning the output to a ``High level" state when input is in a ``Low Level" state and vice versa. 

{To the best of our knowledge, this is the first time that the "Not" task with temporal stimuli is considered. In the parameterization that we designed, when the network receives a stimulus, it must remain with the activity at zero level, while if it does not receive it, after the response interval time that we have predefined in the training set, it must respond. This implementation of the task can be interpreted as we are teaching the network to measure an interval of time.} 

{For the training to be successful in this task, it is the only one where it is necessary to consider a bias term, that is, an initial value for the activity $h_i(t)$ of the units other than zero. In the others, it is not necessary to consider it for the training success. The networks trained for all other tasks from this study have a bias equal to zero. This can be easily changed in the framework because it is just a parameter in the code for the network topology. "Not" is the only task that requires a bias term so that it can be learned by the network with the same conditions described before.}

\begin{figure}[htb!]
\begin{center}
\hspace{-1cm} \textbf{a)}\includegraphics[totalheight=3.35cm]{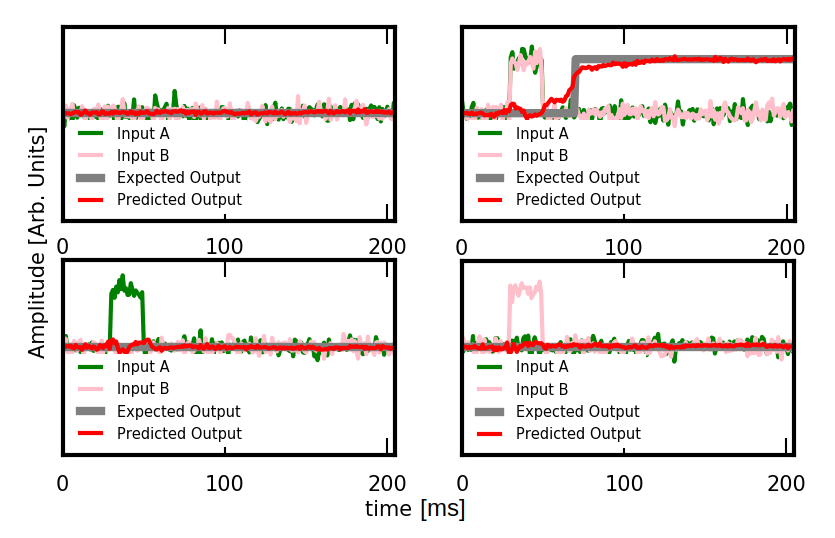}
\textbf{b)}
\includegraphics[totalheight=3.35cm]{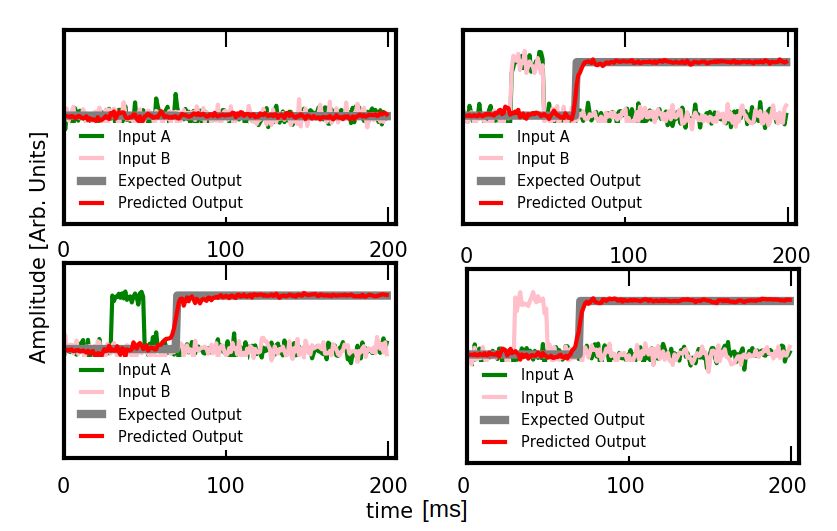}
\vspace{0.5cm}
\hspace{-0.8cm}\textbf{c)}\includegraphics[totalheight=3.25cm]{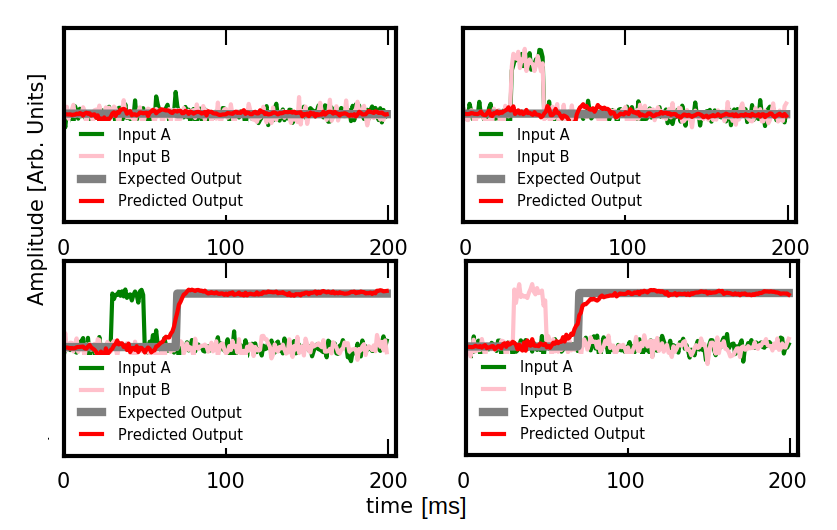}
\textbf{d)} \includegraphics[totalheight=1.8cm]{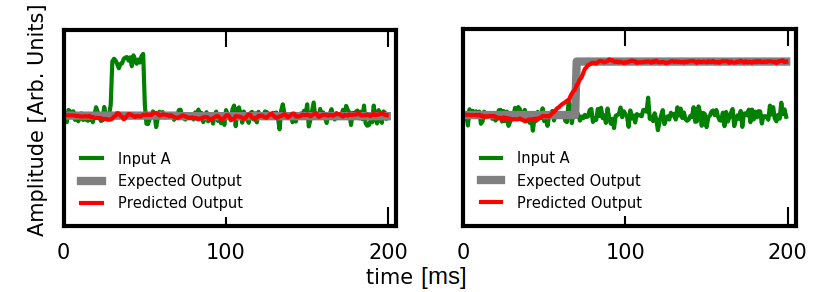}
\caption{Four trained neural networks responses for 6 testing samples for \textbf{a)} AND operation between two stimuli,\textbf{b)} OR operation, \textbf{c)} XOR operation and \textbf{d)} NOT operation applied to the input. Time is in ms and amplitude in arbitrary units.}
\label{fig_03}
\end{center}
\end{figure}

In Figure panel \textbf{d} of Figure \ref{fig_03}, we show the state of the output compared with the input. {In Section \ref{dyn}, we show how the activity looks for an example of a network trained on this task and briefly discuss it.}

Figure \ref{fig_03} shows the temporal response for four neuronal networks, each trained to perform the Boolean operations indicated for each panel when receiving the stimuli at the inputs: AND, OR, NOT, and XOR. The green and pink lines show, in each case, the time series of each input, the thick black line the target output, and the red line the status of the output.

\subsubsection{Flip-Flop (1-bit memory storage)}

In this study, we trained a network with two inputs with different functions. One works as a ``Set" signal (S-input), and the other is a ``Reset" signal (R-input). If the network receives a pulse in the S-input, then the output turns to High. If the network receives a stimulus in the R-input, then the output turns to Low. Two consecutive pulses to the same input do not change the output state. Table \ref{tabla_flipflop} summarizes the rule learned by the network. Time series are 400 ms in length to show different changes in the inputs during the same time-lapse. 

{For this task, we considered the following parametrization: the training data are time series that could have square pulses in the inputs S or R separated by a certain fixed distance, with the condition that the signals in both channels do not overlap. The occurrence of a pulse in one input or another, or its non-occurrence, are random. The response of the output is 20 ms delayed with respect to the input. The time interval between consecutive pulses is not fixed.}

{The Flip Flop task has been previously studied in \cite{DBLP:journals/neco/SussilloB13,SUSSILLO2014156} and many others, but in those works with a task related to a 3-bit register called a 3-bit Flip Flop. Also, this task was studied in Maheshwanarathan et al \cite{maheswaranathan2019universality}. This task was selected because it is not a simple binary decision.}

\begin{figure}[htb!]
\begin{center}
\includegraphics[totalheight=8.5cm]{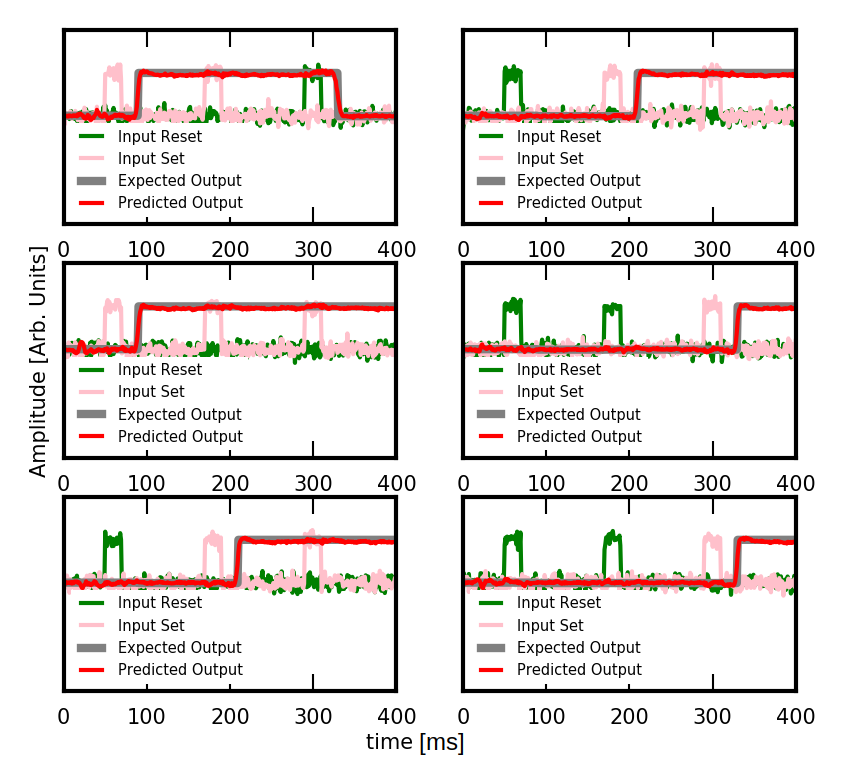}
\caption{{Response of a trained neural network for six random testing samples for the Flip Flop task. S-Signal is shown in pink, the R-Signal is shown in green. Grey thick line is the Target output, and the red is the Network response. The output state depends on the state of the set and resets signals, regardless of the length of the time series considered.}}
\label{fig_07}
\end{center}
\end{figure}

 In Figure \ref{fig_07}, the temporal response of a neural network trained to perform the “Flip Flop” task with its Set and Reset inputs is shown. Each panel shows six possible random time series from the testing data set. The green and pink lines show the time series of each input, the thick black line is the target output, and the red line is the status of the output.
The training data set consists of trains of pulses at the S-input and R-input with noise, and a target output according to Table \ref{tabla_flipflop}. We successfully obtained a set of neural networks capable of performing the Flip-Flop task.

\begin{table}[h!]
\centering
\begin{tabular}{|c|c|c|}
\hline
\textbf{Set} & \textbf{ Reset} & \textbf{Output state} \\ \hline
0                & 0                & $Q_N$           \\ \hline
0                & 1                & 0               \\ \hline
1                & 0                & 1               \\ \hline
1                & 1                & $X$               \\ \hline
\end{tabular}
\caption{Flip Flop task table. $Q_N$ means that the output remains at the previous state. $X$ means that the state is forbidden for the data set.}
\label{tabla_flipflop}
\end{table}

\subsubsection{Finite-duration oscillator}\label{sub-osc}

\begin{figure}[htb!]
\begin{center}

\includegraphics[totalheight=3.5cm]{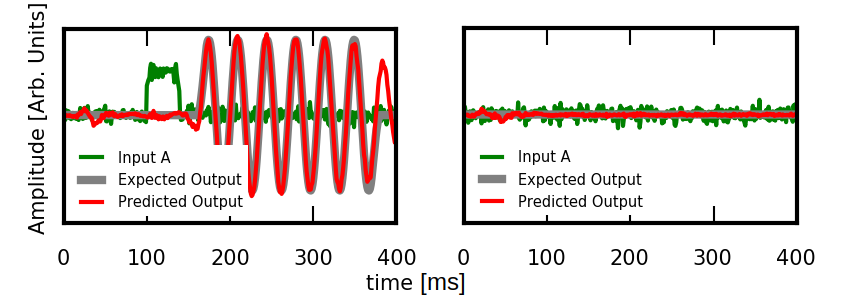}
\caption{A trained neural network response of the output to the testing samples for the ``finite-duration oscillator" task. The network does not spontaneously oscillate unless there is a pulse in the input. }
\label{fig_08}
\end{center}
\end{figure}

In this task, we trained a network to obtain in the output an oscillation with a frequency of $30$ $Hz$, 20 ms after a pulse in the input.  When the network receives the stimulus, the output behaves as it is shown in Figure \ref{fig_08}. If the network has no stimulus, the output remains at the ``LOW" state. Once again, we successfully trained a set of neural networks that perform this task.

Figure \ref{fig_08} shows the temporal response of a neural network trained to oscillate when receiving a stimulus at the input. Upon receiving a pulse at the input, the output must respond with oscillation for a certain time. In all Figures (from 2 to 4), the networks have 50 units and 20 ms for the response time.

{This simple task is interesting, on the one hand, because here the network has to learn to reproduce a pattern when it receives a contextual signal, different from the other tasks that we have considered, which are decision-making tasks. It is also interesting because the activity of the units always results in oscillations, as we discuss in more detail in Section \ref{dyn}. We also used this task in such section to discuss differences related to what happens with noise on the input and in the absence of it.}

\subsection{Network initialization studies} \label{net-init}

{In this section we show the} different properties of the network activity and connectivity based on population analysis employing PCA and the estimation of the eigenvalues of the recurrent weight matrix. 

\begin{figure}[htb!]
\begin{center}
\includegraphics[totalheight=5.25cm]{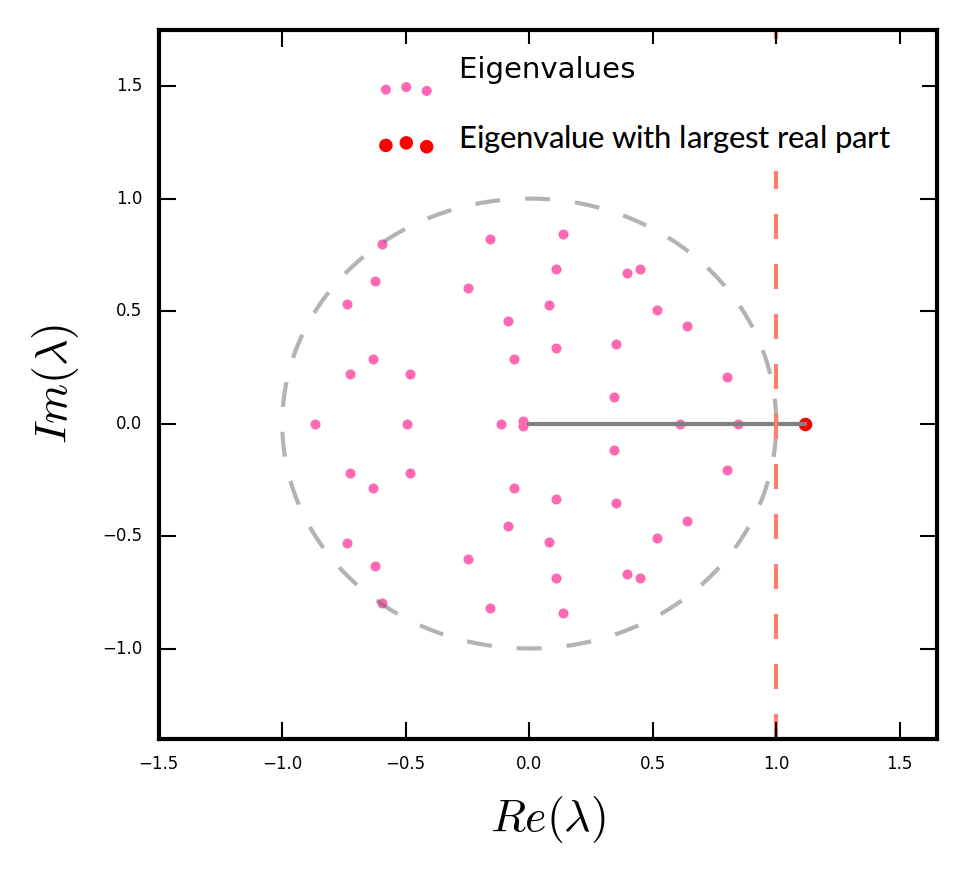}
\includegraphics[totalheight=5.25cm]{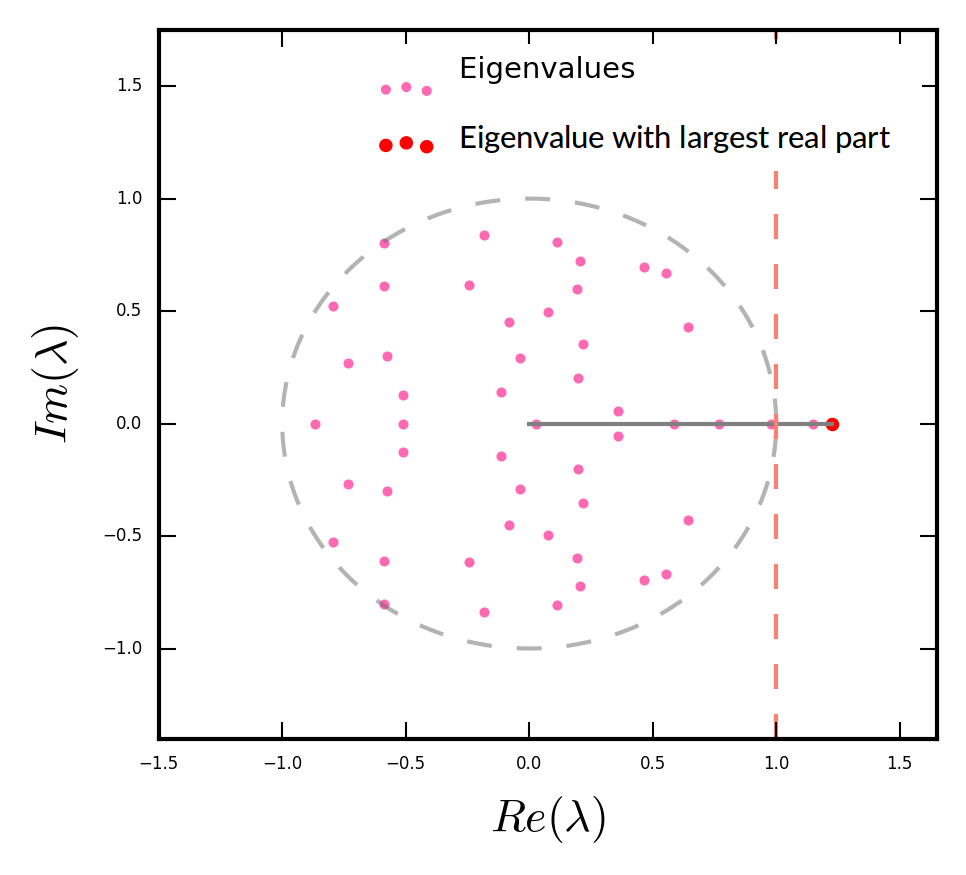}
\includegraphics[totalheight=5.25cm]{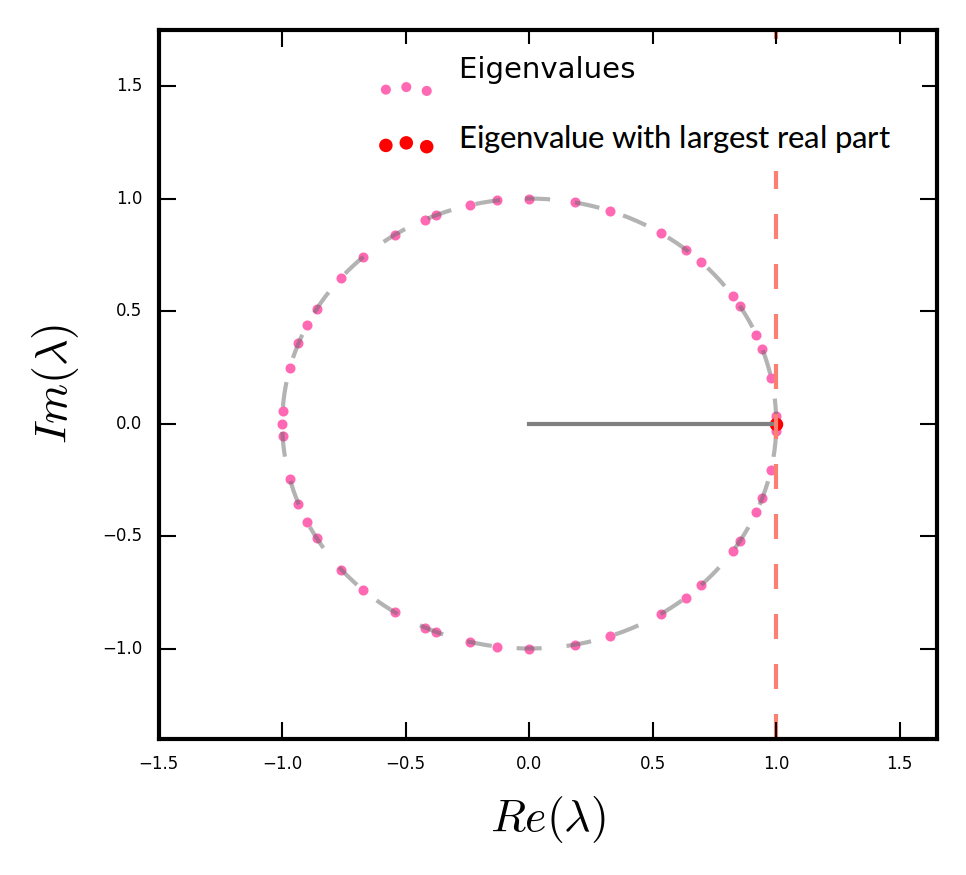}
\includegraphics[totalheight=5.25cm]{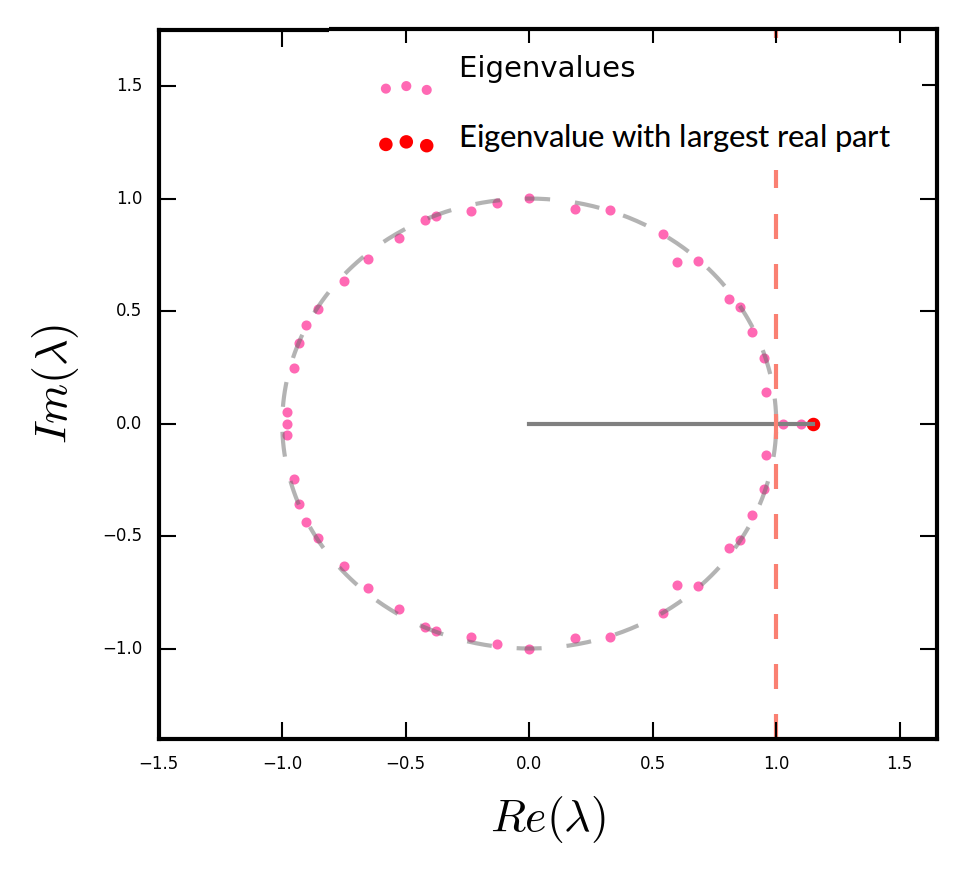}
\caption{Eigenvalue spectrum: Left. Eigenvalue distribution for a Neural network with initial configuration random normal (upper panel) and orthogonal (bottom panel). Right. Eigenvalue distribution of the $\mathbf{W^{Rec}}$ matrix post-training AND ($\#ID14$ for orthogonal condition, $\#ID07$ for random normal).}
\label{fig_08_A}
\end{center}
\end{figure}

We started by training a set of networks to perform each of the considered tasks described in the previous section. Each network was created by randomly choosing the connectivity strengths $\mathbf{W^{Rec}}$ from a normal distribution with zero mean and variance $1/N$, as described in Section \ref{model}. We considered two cases: orthogonal matrices and non-orthogonal matrices (20 matrices in each case). We studied the eigenvalue spectrum of the recurrent matrix $\mathbf{W^{Rec}}$. The eigenvalues of two sample matrices are shown in the upper and lower left panels of Figure \ref{fig_08_A}.

The non-orthogonal matrix, previous to the training (upper left panel), shows a distribution consistent with the random matrix theory of Girko's circle law \cite{doi:10.1137/1129095}, which states that, for large N, the majority of eigenvalues of an $N \times N$ asymmetric random matrix lie uniformly within the unit circle in the complex plane and the fraction of eigenvalues lying outside the circle vanishes in the limit $N_{units}\rightarrow  \infty$, when the elements are chosen from a distribution with zero mean and variance $\frac{1}{N}$.

In the orthogonal matrix, the eigenvalues lie at the border of the circle (bottom left panel of Figure \ref{fig_08_A}). 
As a result of the training, some eigenvalues are pushed out of the circle. In the case of the networks shown in Figure \ref{fig_08_A}, both final configurations correspond to fixed-point configurations with one eigenvalue with real part greater than one and zero imaginary part, and the rest of the eigenvalues scattered within the unit circle. Here we show a comparison between the initial state (left panels) and post-training (right panels) of Figure \ref{fig_08_A}. {We have found similar configurations of the eigenvalue distribution for all the trained networks.}

We found in our simulations that this behaviour is consistent with estimates made previously in \cite{10.1371/journal.pcbi.1006309, PhysRevE.88.042824, PhysRevResearch.2.013111}, even when our networks are trained (i.e. non-random) and input-driven. The system will either show nontrivial stationary solutions or oscillations depending on the value of the eigenvalue with the largest real part of the connectivity matrix as in \cite{PhysRevE.88.042824}. A more profound explanation is beyond the scope of the present manuscript.

For each of the tasks and the two conditions (orthogonal and non-orthogonal), we estimated the rate of networks that successfully passed the training. The results are shown in Table \ref{Tabla-rate}. {We measure the success rate as the number of trained networks that successfully reproduce each task, with respect to the total number of networks that we trained, considering a fixed number of epochs, which is 20 as defined in Section \ref{protocol}.}

The orthogonal condition slightly improves the success rate for each task. This is consistent with studies previously conducted by \cite{Vorontsov2017OnOA}. A possible explanation for the success rate differences between the two possible initial conditions is that at the training stage is ``easier’’ to pull out the eigenvalues when they are placed on the edge of the circle (orthogonal condition) than when they are scattered within the unit circle (non-orthogonal condition). 

{Our results are consistent with  \cite{Vorontsov2017OnOA} in the sense that, for a simple task, and not complex architectures, such as the LSTM extensively analyzed in ML, we observe also improvements in the training performance when initializing matrices with the orthogonal condition. Orthogonal initialization shapes the position of the non-dominant eigenvalues, but we observed that it has no additional effect on the dynamics of the obtained realizations. The only effect is in the training performance, not in the activity after training.}

The time reproduction task shows a perfect training rate (100\%) with both initializations. We think that this is because this is the simplest task to be learned for the network. This is also why we chose this task for our scaling studies in the following sections.

\begin{table}[h!]
\begin{center}
\begin{tabular}{|c|c|c|}
\hline
Task        & Initial orthogonal  & Initial Rand Normal\\
\hline
{AND}         & 85\%                     & 65\%              \\\hline
{OR}          & 90\%                     & 80\%              \\\hline
{XOR}         & 90\%                     & 55\%              \\\hline
{NOT}         & 90\%                     & 45\%              \\\hline
Flip Flop (1-bit memory storage)  & 95\%                     & 65\%              \\\hline
Oscillatory & 90\%                     & 65\%              \\\hline
Time reproduction  & 100\%                    & 100\%              \\        
\hline
\end{tabular}
\caption{The success rate for the training of 20 networks for orthogonal initial condition compared with the random normal initial condition.}
\label{Tabla-rate}
\end{center}
\end{table}

From Figure \ref{fig_08_A}, it is interesting to note that there are a few eigenvalues that dominate the behaviour of the states that result in the task for which the network was trained.
 
The correlation between eigenvalue spectra and the dynamics of neural networks has also been studied in \cite{doi:10.1162/neco.2009.12-07-671}, relating the design of networks with memory face with the eigenvalues outside the circle in the complex plane. 

\subsection{Network dynamics}\label{dyn}

We plot the components $h_i(t)$ from the $\mathbf{H(t)}$ from Equation \ref{eq-03}, this is the temporal evolution of all recurrent units. We applied Principal Component Analysis (PCA) to this set in each case. {This means that we performed a decomposition into Principal Components with the entire set of activity for the output's units {$h_i(t)$} for each different relevant condition. It was done for each input combination and task, using the method from the scientific library Scickit Learn \cite{scikit-learn}. These are Python open-source libraries based on Numpy that allow us to perform dimensionality reduction, feature extraction, and normalization, among other methods for predictive data analysis. The behaviour of the system was plotted into the 3 axes of greatest variance. }

{Our approach involves analyzing the response to noiseless stimuli to depict the undisturbed trajectory and construct a geometrical representation of the phase space}. All networks were trained with noisy input signals, as previously described. 

\begin{figure}[htb!]
\begin{center}
\includegraphics[totalheight=5.5cm]{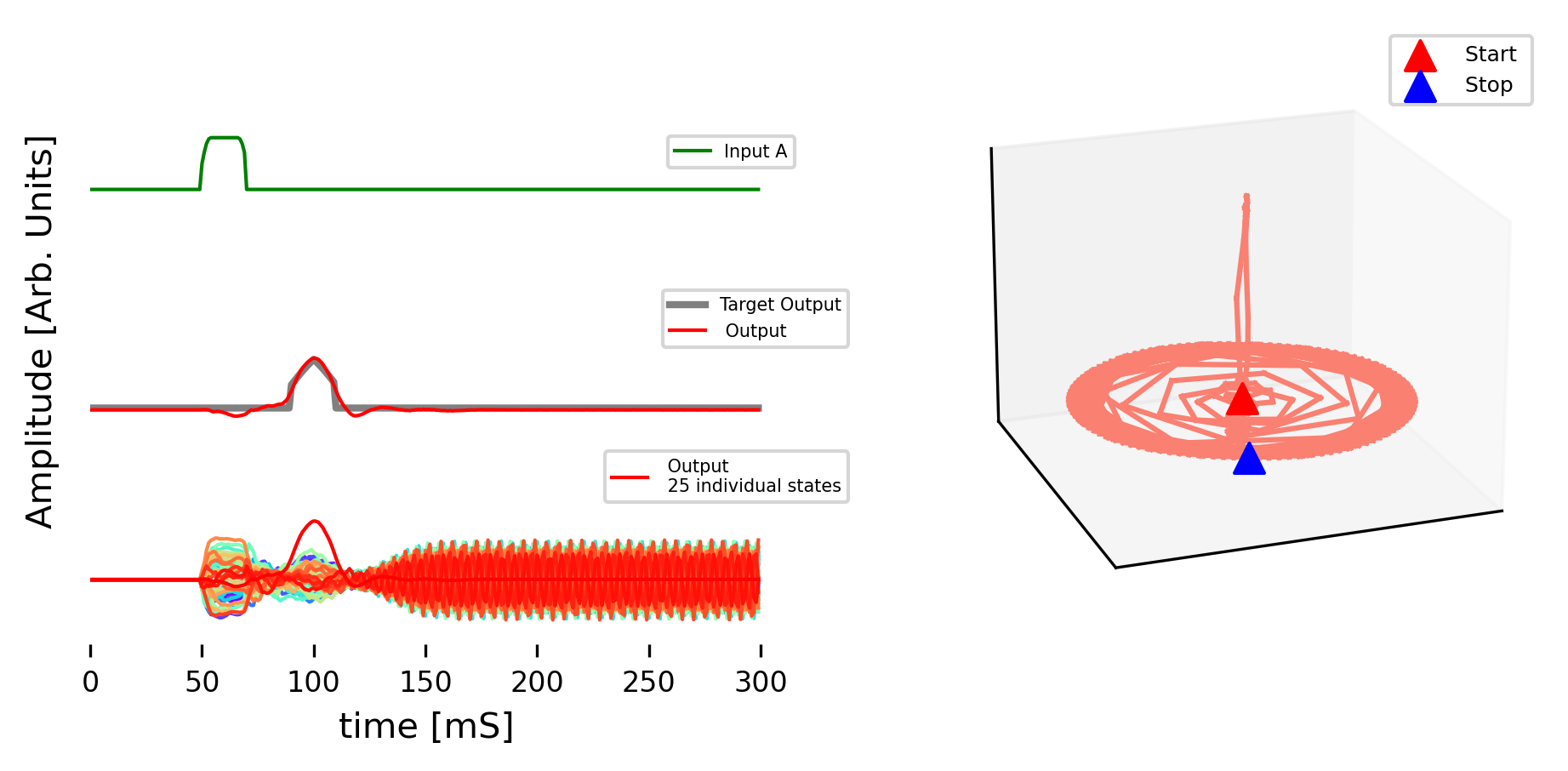}
\caption{Neural network $\#ID03$ response in the time reproduction task.}
\label{fig_08_c}
\end{center}
\end{figure}

{Let us first consider the simpler task proposed in section \ref{tasks-1}. In this case, for all the realizations obtained and initial conditions, we observe that after receiving the input stimulus, the activity of the units combines to give zero the value at readout for the duration of the pulse. After the delay time, the signals combine to give rise to the output pulse, and then the activity is oscillatory. Different trained networks present variations in frequencies and amplitudes of oscillations, but they all converge to the same general behaviour as shown in Figure \ref{fig_08_c}.}

Now we consider networks trained in the AND task. We want to show the behaviour of the recurrent units. We studied the response to the stimuli corresponding to the four different input configurations (Table \ref{tabla_and}).

Figure \ref{fig_09} shows the behaviour of a trained network (labelled as $\#ID14$ in Supplementary Materials) when the stimulus is applied. The left side of the Figure shows the output and inputs signals vs. time in the upper panel and some recurrent units $h_i(t)$. {The right panel shows the trajectory in the state space of the first three Principal Components. The results are shown in Figure \ref{fig_09} with the ``Low-Low", ``High-High", ``Low-High", and ``High-Low" combinations (top to bottom)}. All four cases are represented by stable fixed points. When no pulse is applied to the inputs, the activity of every recurrent unit is zero (Figure \ref{fig_09} (a)). When both inputs are pulsed, the recurrent units are perturbed, and then the system migrates to a fixed point that is different from zero (Figure \ref{fig_09} (b)). When any single input is pulsed (Figure \ref{fig_09} (c) and (d)), the network dynamics converge to fixed points that are different from the previous two and are also different from each other. It can be seen that the recurrent units change their activity to reproduce the learned behaviour (HIGH output) with a different final internal state that depends on which input was activated. 

\begin{figure}[!tbph]
\begin{center}
\hspace*{-1cm}\includegraphics[totalheight=4.5cm]{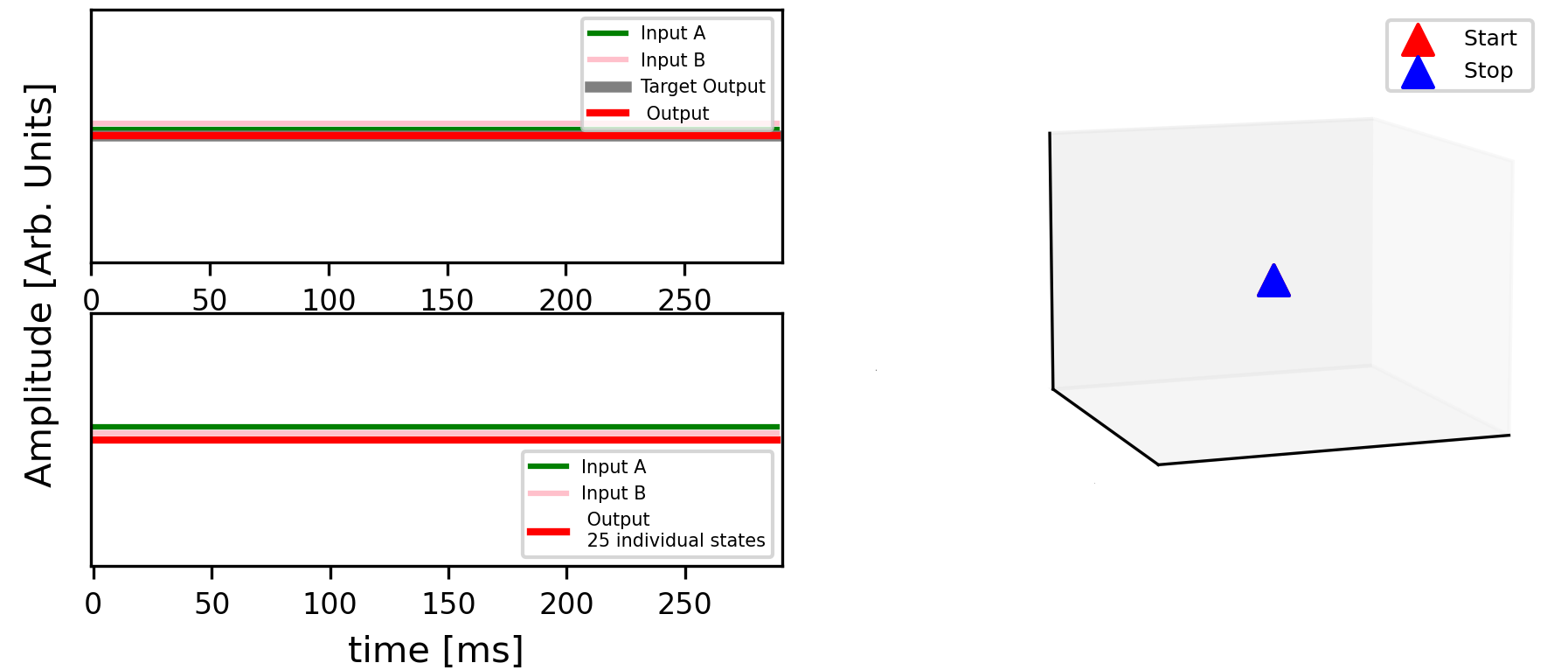}
\hspace*{-1cm}\includegraphics[totalheight=4.5cm]{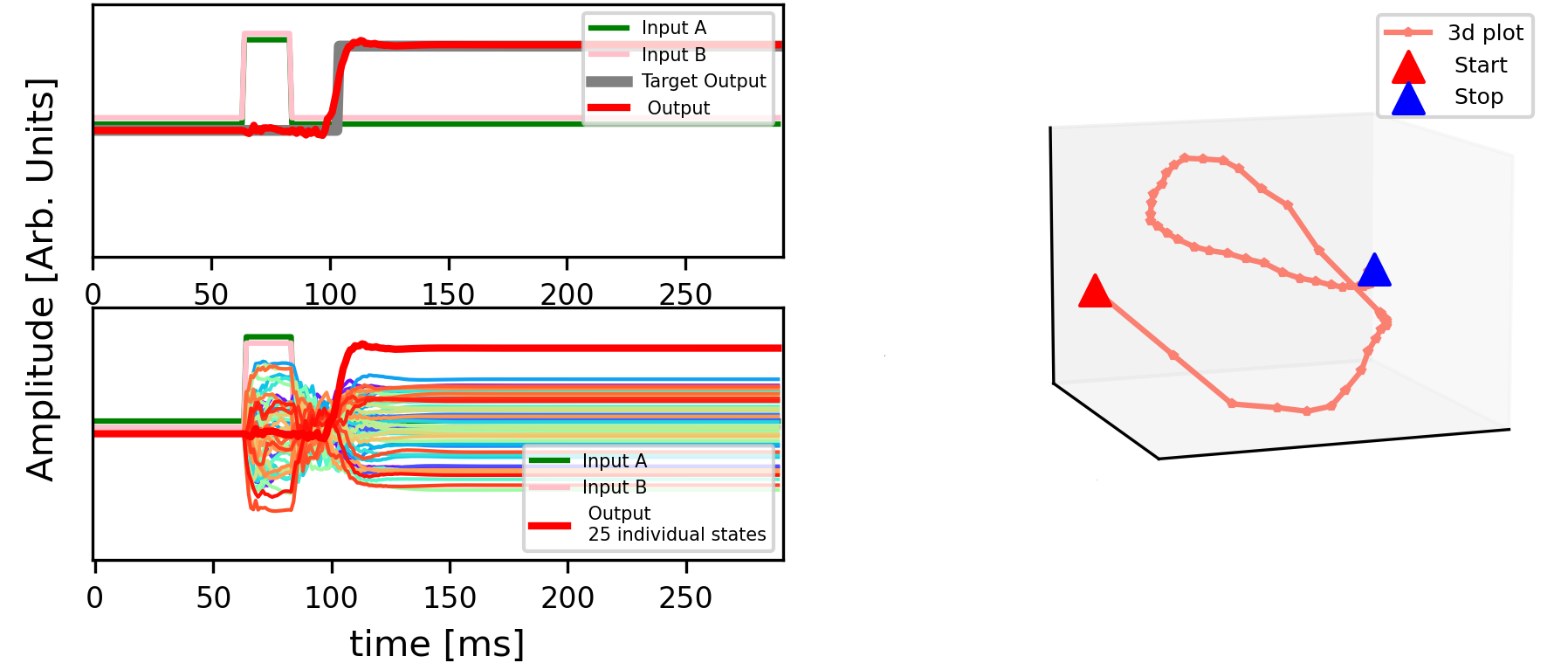}
\hspace*{-1cm}\includegraphics[totalheight=4.5cm]{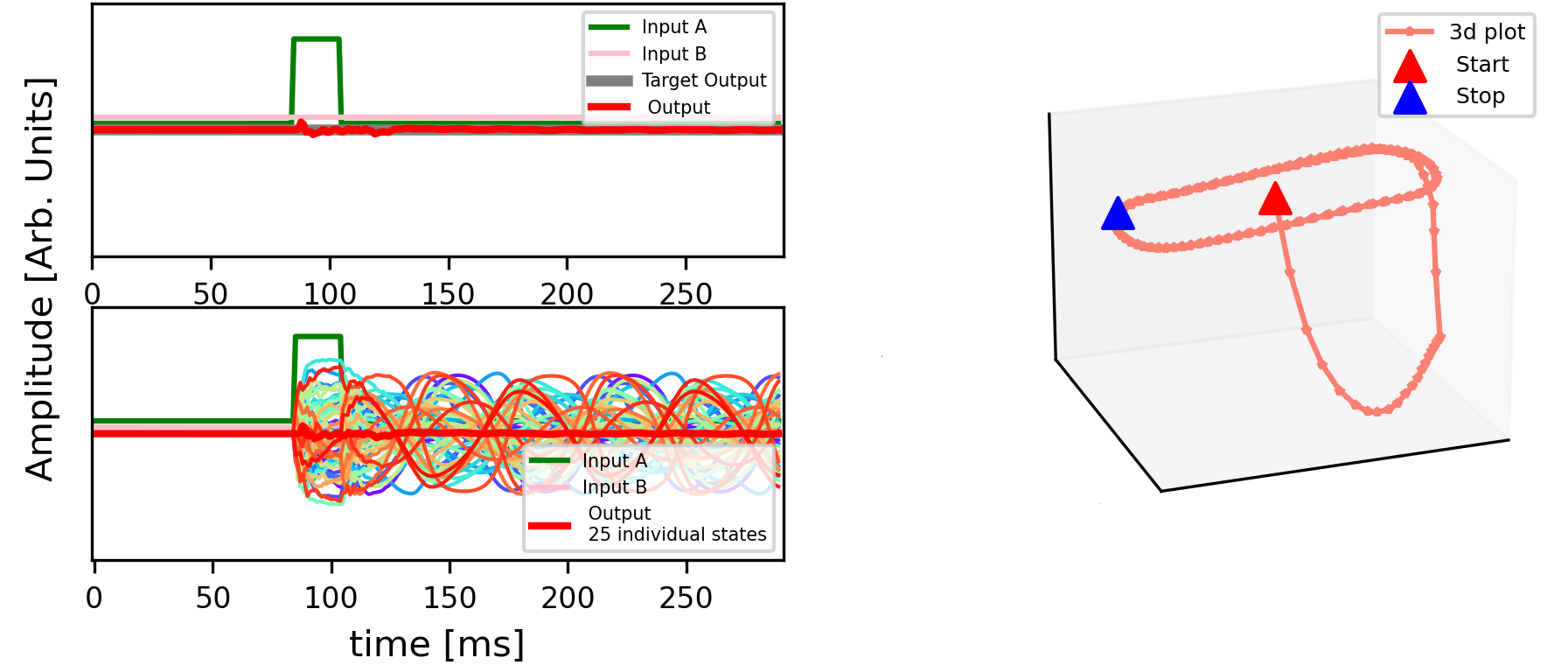}
\hspace*{-1cm}\includegraphics[totalheight=4.5cm]{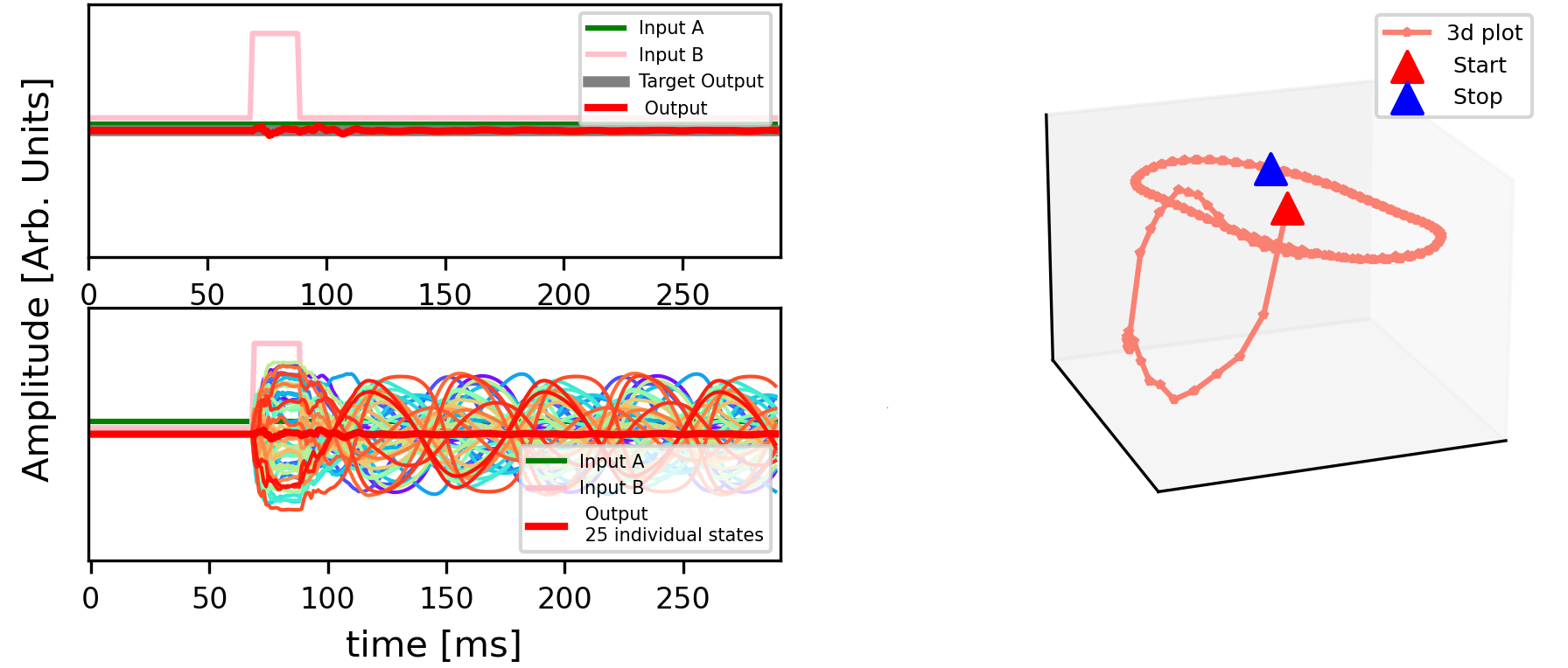}
\caption{Left. Neural network $\#ID04$ response of the output in the AND task for each possible input combination (a-d). Right: PCA analysis for the same dataset. The top left panel shows inputs, the target and output. The bottom left panel shows 25 individual $h_i(t)$ states.}
\label{fig_11}
\end{center}
\end{figure}

We repeated this analysis for all trained networks and found that the way to achieve the desired trained rule is not unique, which is consistent with \cite{maheswaranathan2019universality}. We identified different dynamical regimes for learning the same rule by different networks. We show in Figure \ref{fig_11} the results for a different network trained in the AND task. The High output state (corresponding to ``Low-Low" and ``High-High" stimuli for the inputs, Figure \ref{fig_11}) is solved with two different stable fixed points as before, yet the Low output state (``Low-High" and ``High-Low" stimuli for the inputs, Figure \ref{fig_11}) is represented by two different limit cycles (i.e. an oscillation in the activity of the recurrent units). In this case, the  $\mathbf{W^{Rec}}$ matrix has one real and two complex conjugated leading eigenvalues.

From these results, it is clear that the same task can be performed with different internal configurations. In the case of the first network ($\#ID14$, Figure \ref{fig_09}), the task is solved by converging to stable fixed point recurrent states. In this case, the distribution of eigenvalues of the trained matrix has pure real dominant eigenvalues.

On the other hand, in the network identified as $\#ID04$ ({Figure \ref{fig_11}}) the input configurations ``01" and ``10" produce oscillatory recurrent states, while ``00" and ``11" produce stable fixed point recurrent states. permanent. In this case, the leading eigenvalues are complex conjugates (i.e. nonzero imaginary part).

{The same situation occurs for the other configurations when considering the different tasks studied, except for the case of oscillation caused by a stimulus presented in Section \ref{sub-osc}, where the internal state is always oscillatory.}

\begin{figure}[!tbph]
\begin{center}
\hspace*{-1cm}\includegraphics[totalheight=4.5cm]{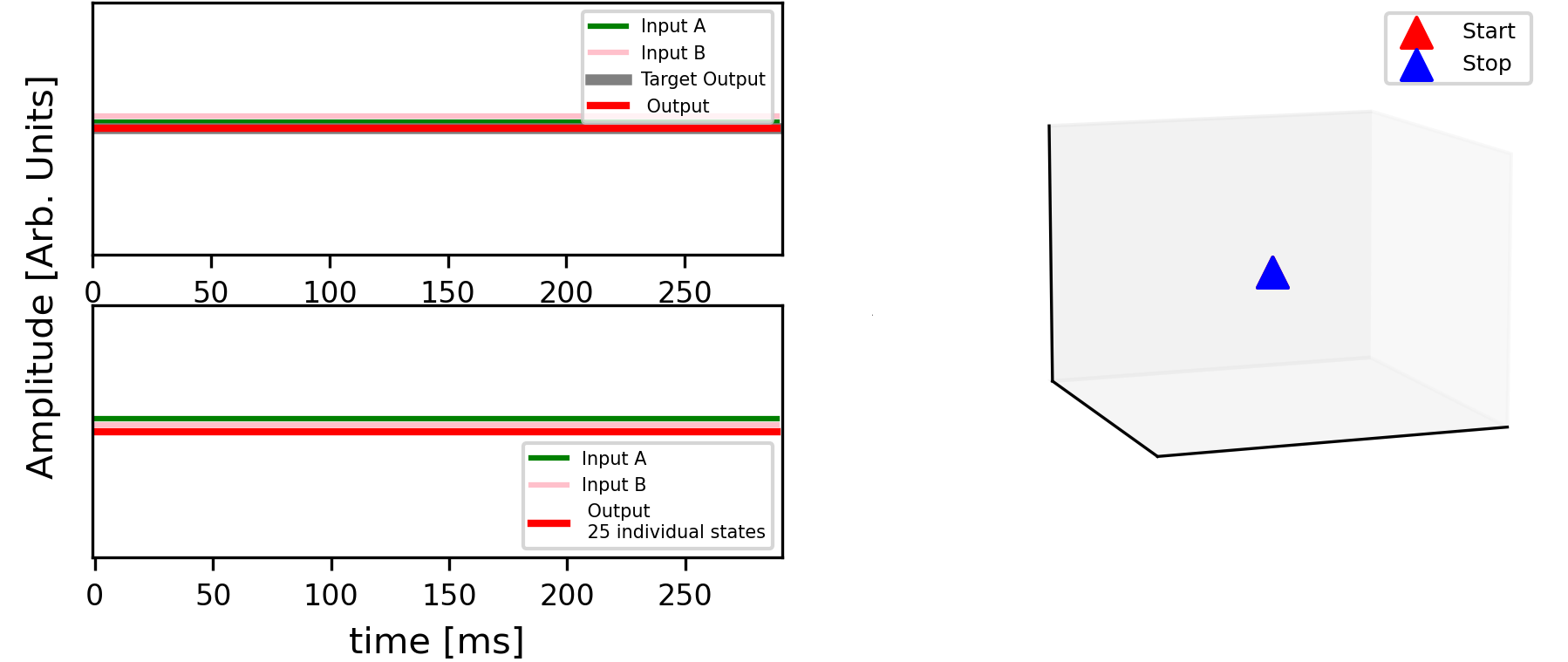}
\hspace*{-1cm}\includegraphics[totalheight=4.5cm]{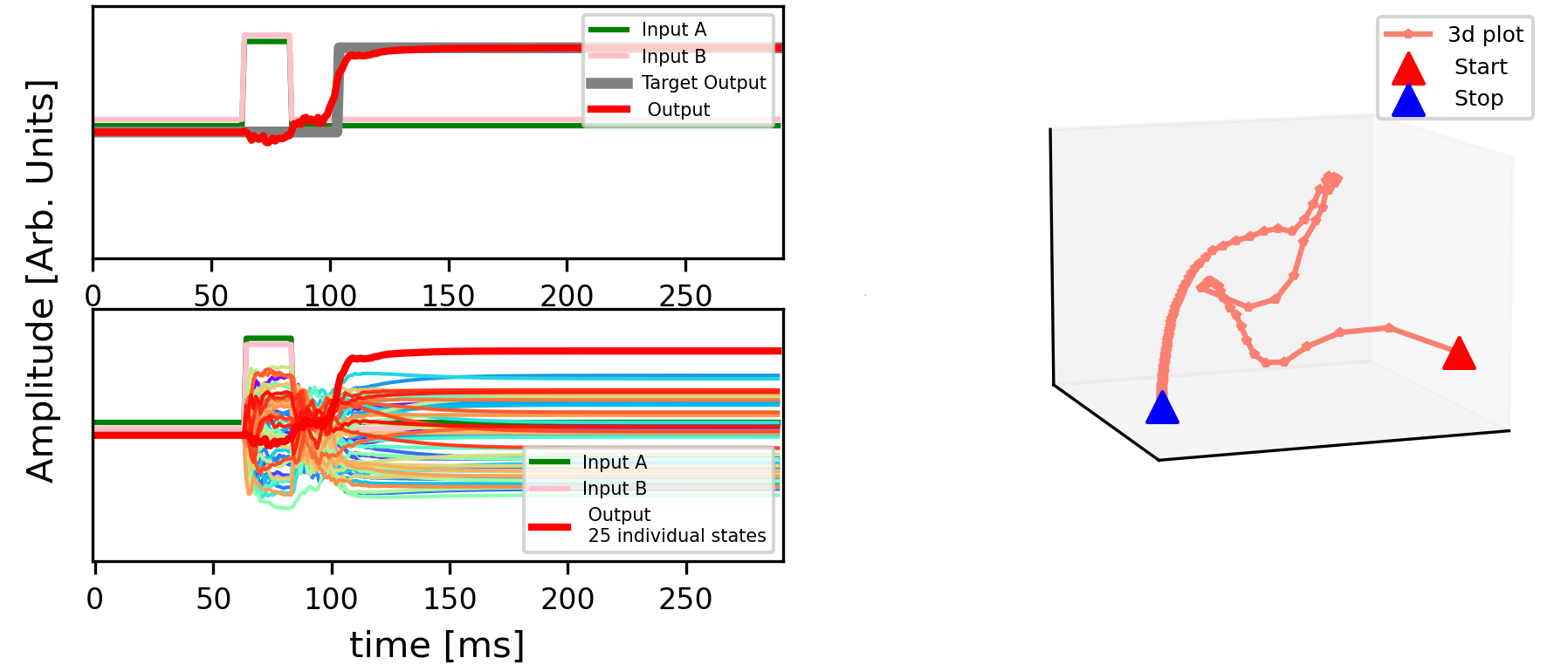}
\hspace*{-1cm}\includegraphics[totalheight=4.5cm]{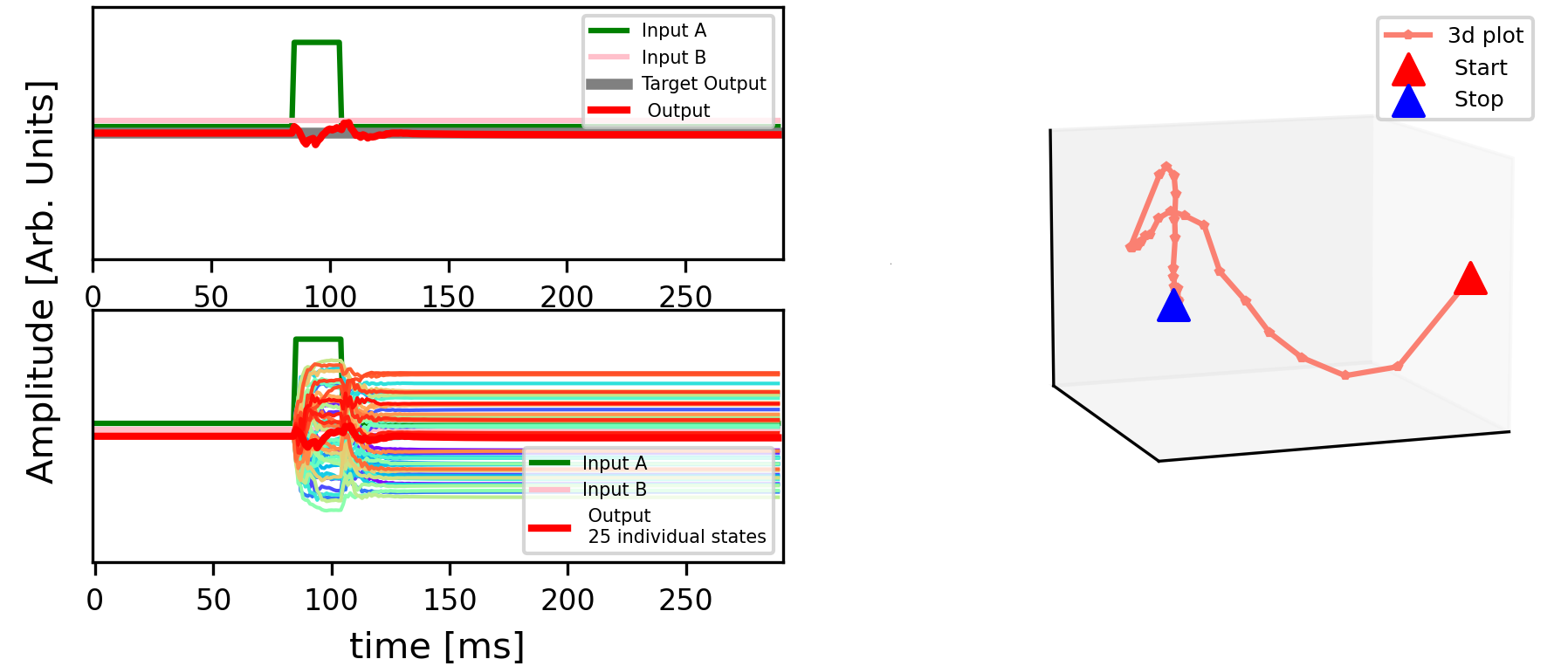}
\hspace*{-1cm}\includegraphics[totalheight=4.5cm]{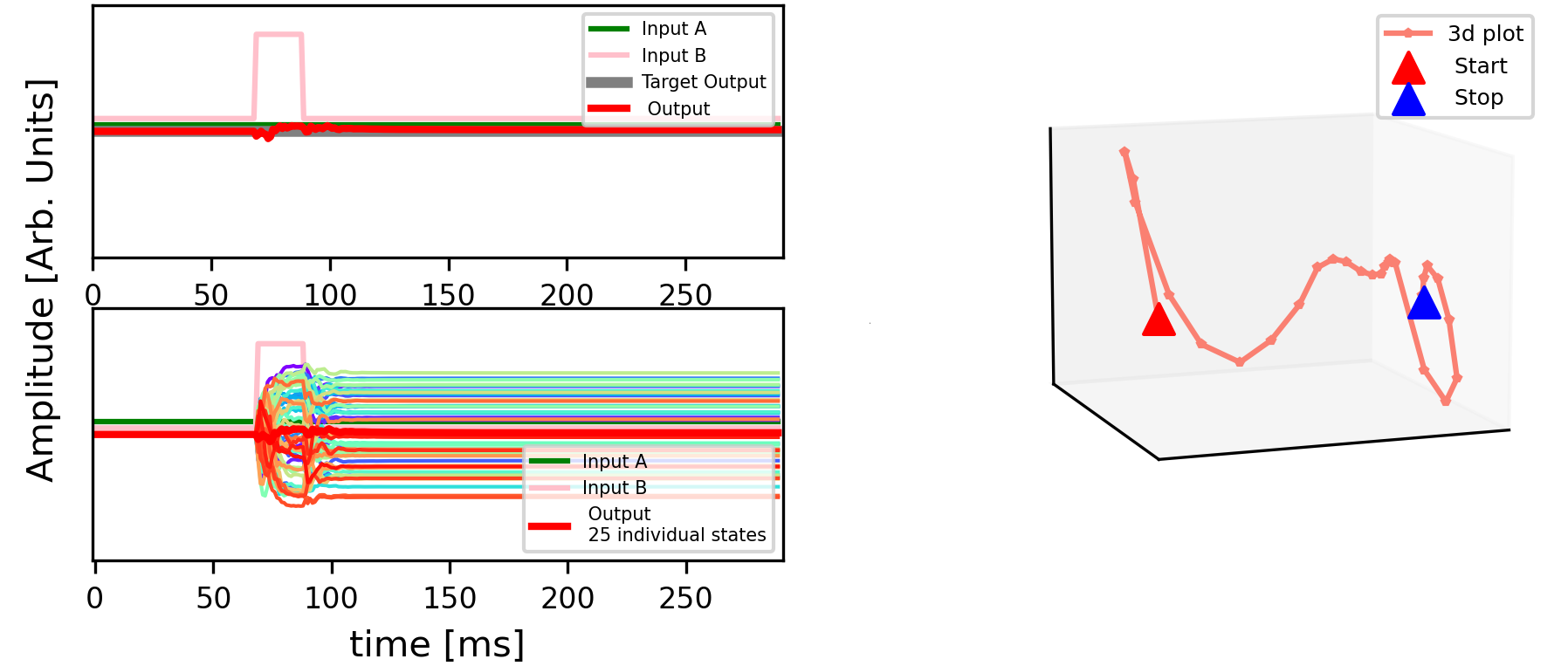}
\caption{Left: Neural network $\#ID14$ response in the AND task for each possible input combination (a-d). Both configurations are represented by stable fixed points. Right: the first three PCs for the same dataset.}
\label{fig_09}
\end{center}
\end{figure}

It is worth noting that we have found the same behaviour in larger networks (500 units), i.e. behaviour driven by a small set of eigenvalues and a variety of dynamical solutions underlying the same learned task (See Supplementary Materials). On the other hand, and perhaps not surprisingly, the finite-duration oscillation task was learned by all networks in the same way---a limit cycle \cite{DBLP:journals/neco/SussilloB13}. 

\begin{figure}[!tbph]
\begin{center}
\hspace*{-1cm}\includegraphics[totalheight=5cm]{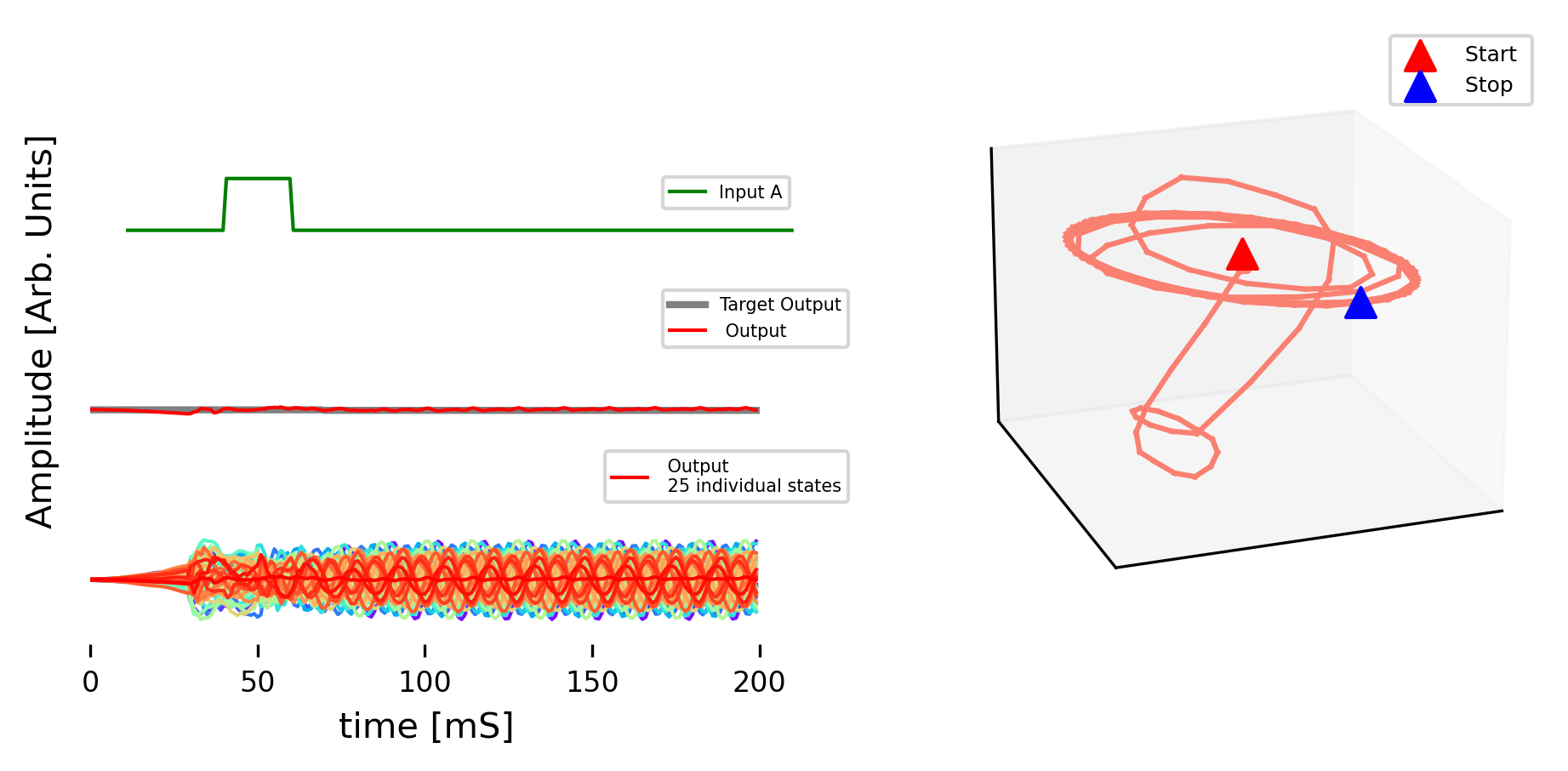}
\hspace*{-1cm}\includegraphics[totalheight=5cm]{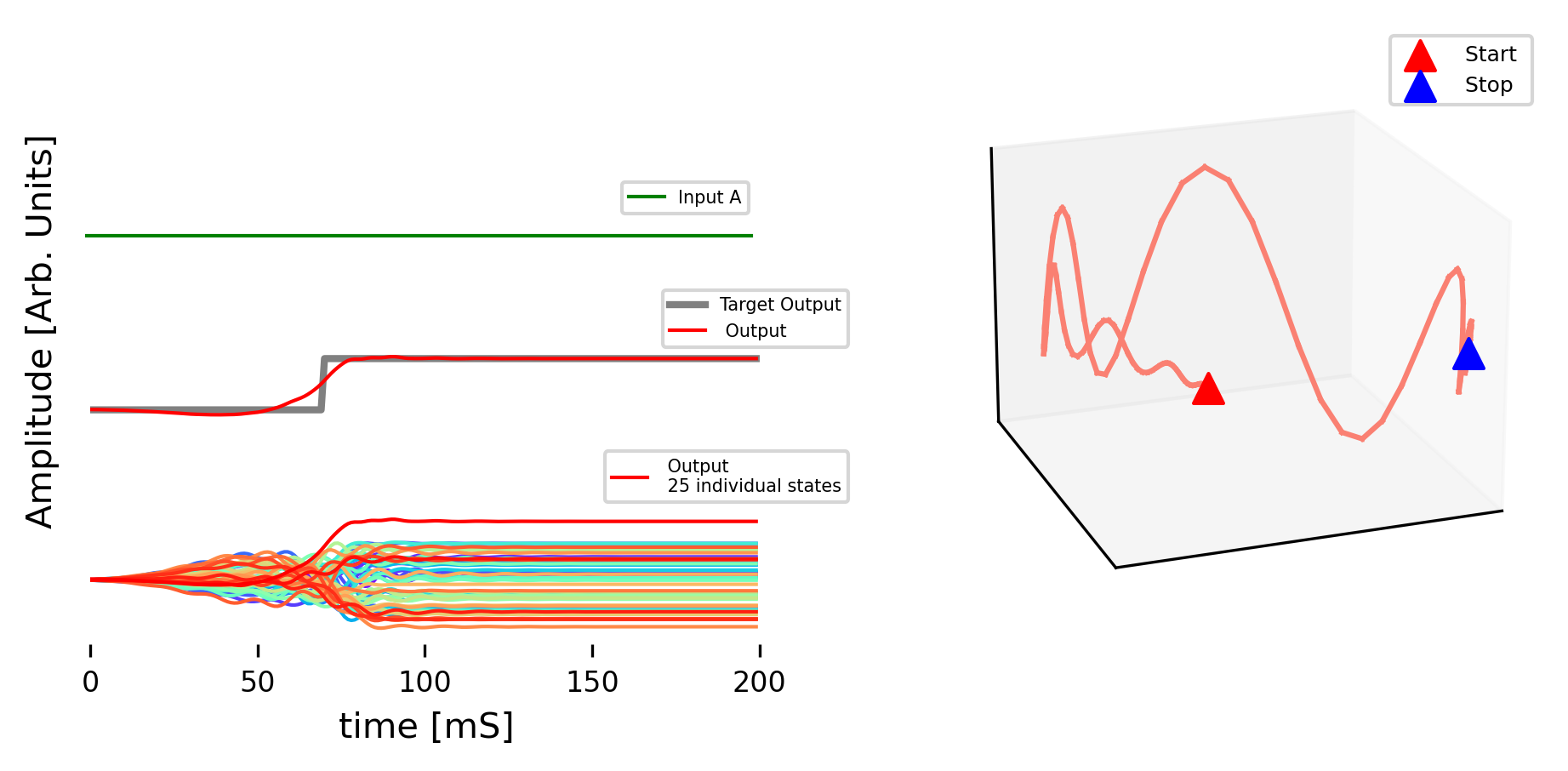}
\caption{Neural network $\#ID17$ response of the output in NOT task for the two possible situations.}
\label{fig_12b}
\end{center}
\end{figure}

{Let us now consider the response of the units of a network trained for the "NOT" task (Figure \ref{fig_12b}) as we have described in Section \ref{not}. It is interesting to note the behaviour of the activity in the absence of a stimulus. In this case, the network must learn the time interval at which it must respond without receiving any stimulus.}

{Different realizations for the same task showed to converge to similar solutions, meaning they tend to fixed points when the input does not receive a stimulus and oscillations when the output must remain at zero. The difference between realizations is that  the oscillations are usually of different frequencies and amplitude.}

\begin{figure}[!tbph]
\begin{center}
\hspace*{-1cm}\includegraphics[totalheight=5.5cm]{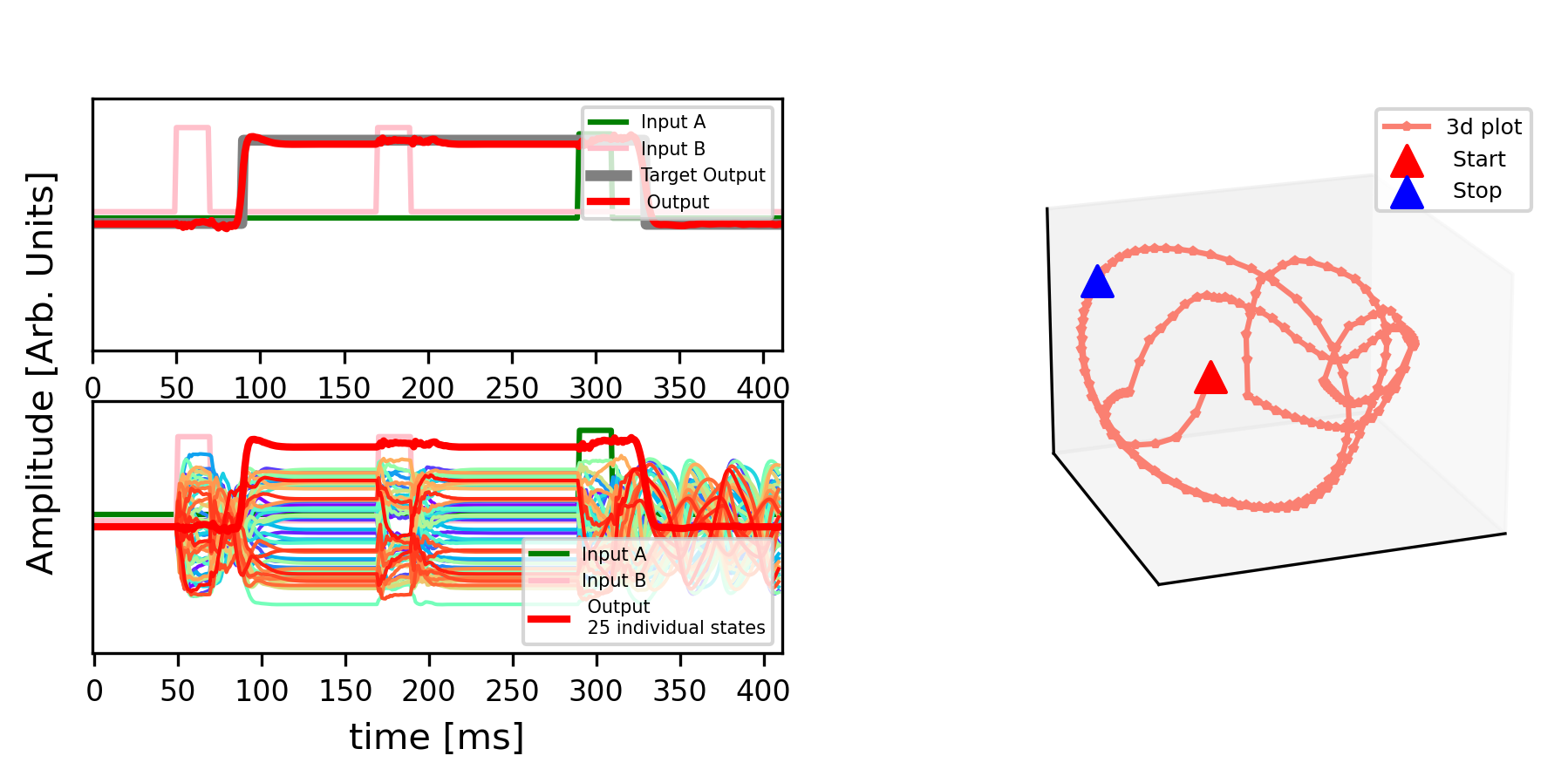}
\hspace*{-1cm}\includegraphics[totalheight=5.5cm]{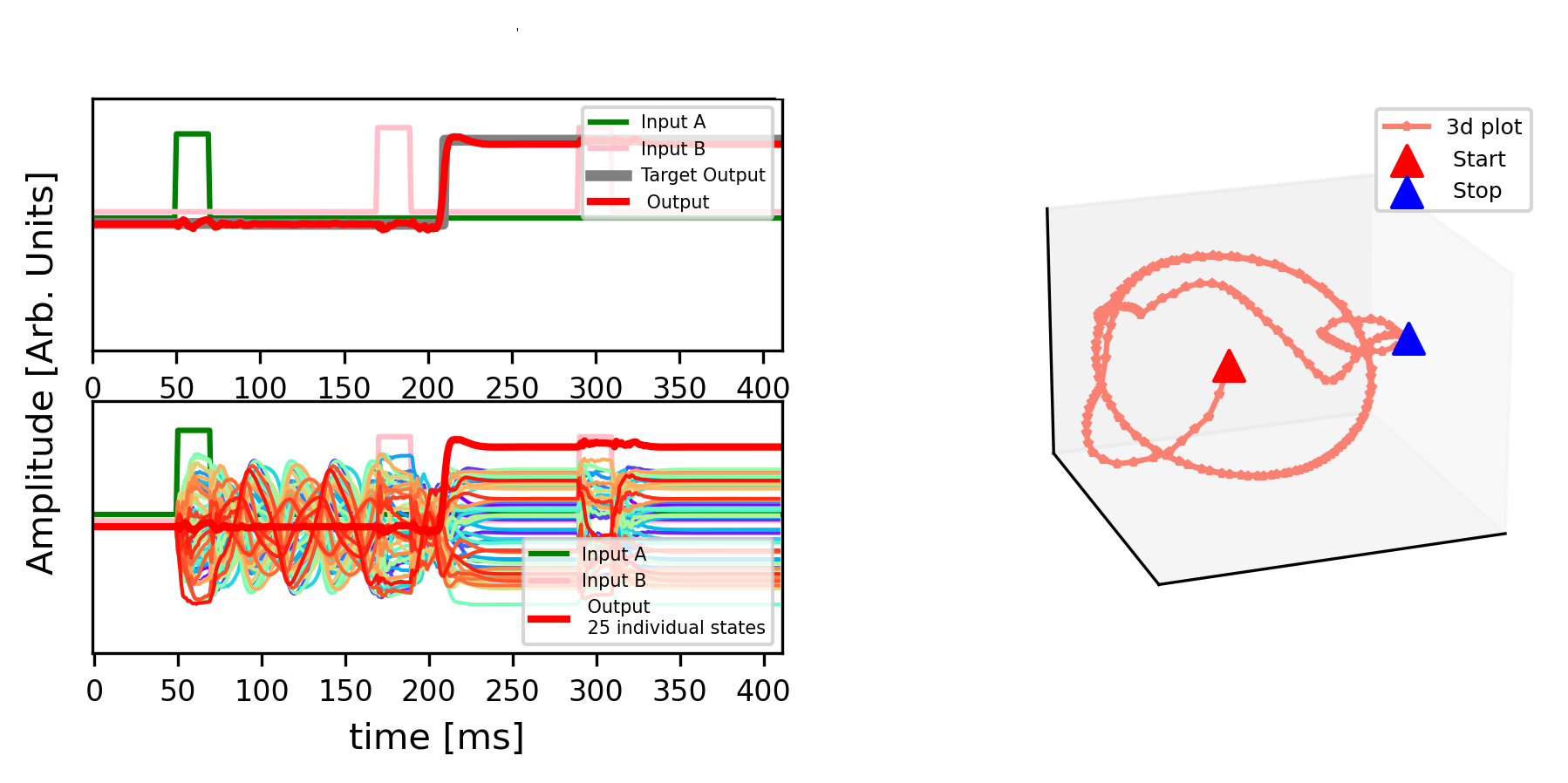}
\includegraphics[totalheight=6cm]{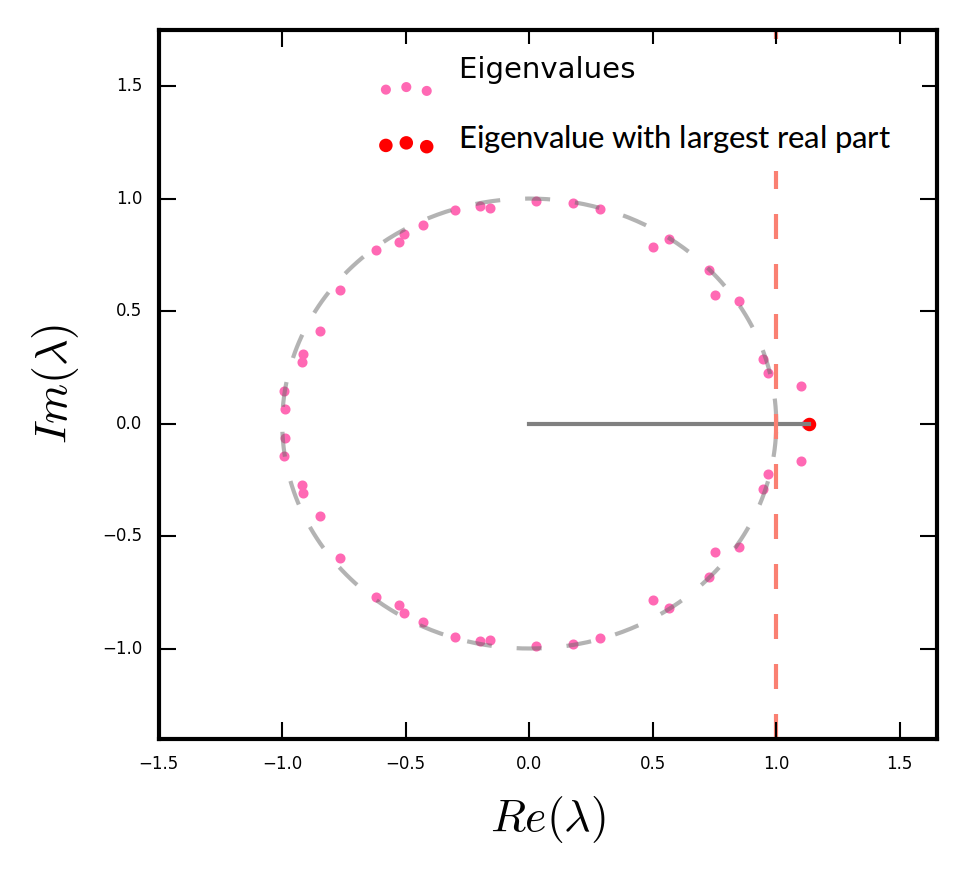}
\caption{Left. Neural network $\#ID06$ trained in the Flip Flop task. Right PCA analysis for the same dataset. The top left panel shows inputs signals, the target and output. The bottom left panel shows 25 individual $h_i(t)$ states. Bottom panel: Eigenvalue spectrum of the $W^{Rec}$ matrix.}
\label{fig_10_b}
\end{center}
\end{figure}

Next, we show a network that learned the Flip-Flop Task in Figure \ref{fig_10_b}. Input A represents the ``Set" Signal, and input B represents the ``Reset". The upper panel shows an example of ``Set’’ followed by ``Reset", and the lower panel ``Reset" followed by ``Set’’. The High output solution is a stable fixed point, while the Low output is a stable limit cycle. It is interesting to note that a reset signal will take the system to a state different from the one that started before stimulation, even if the output must go back to the same value (zero). This behaviour is also ruled by the three leading eigenvalues: one real and a set of complex conjugated, as it is shown in the bottom panel of Figure \ref{fig_10_b}.

{Finally, here we discuss the effect of noise. As mentioned before, all the networks were trained with stimuli containing random noise levels throughout the entire training data set. To visualize the behaviour of the dynamics, we excited them with stimuli with and without noise. In all tasks, the only change observed is the clarity of the trajectory in the reduced dimensional space. Showing the dynamics in response to a noiseless stimulus results in a more clear trajectory in the reduced dimension space for all decision-making tasks and the time reproduction task.
However, when we consider the task of the finite duration oscillator stimuli with and without noise produced a different effect.}

\begin{figure}[!tbph]
\begin{center}
\hspace*{-1cm}\includegraphics[totalheight=5cm]{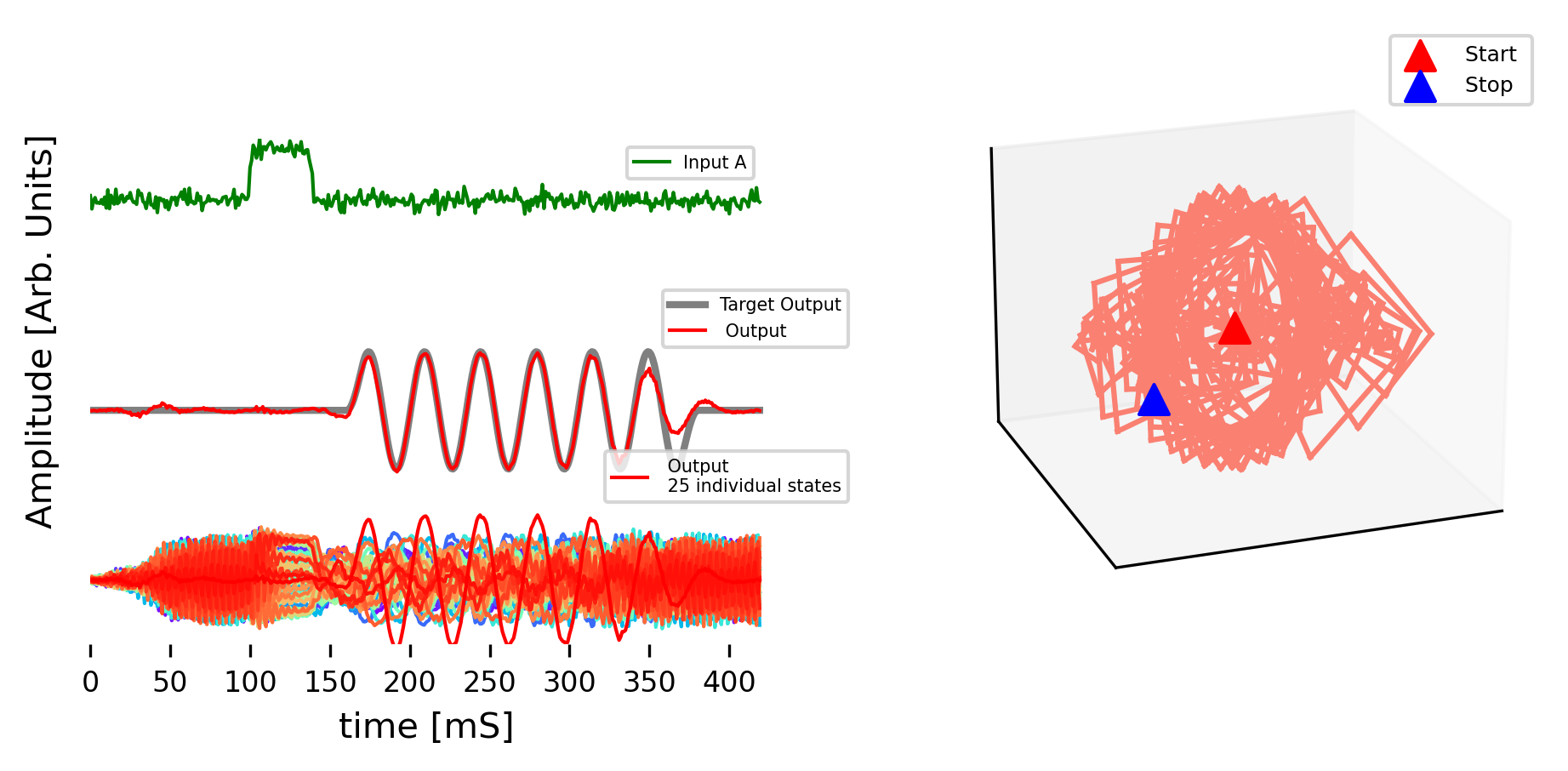}
\hspace*{-1cm}\includegraphics[totalheight=5cm]{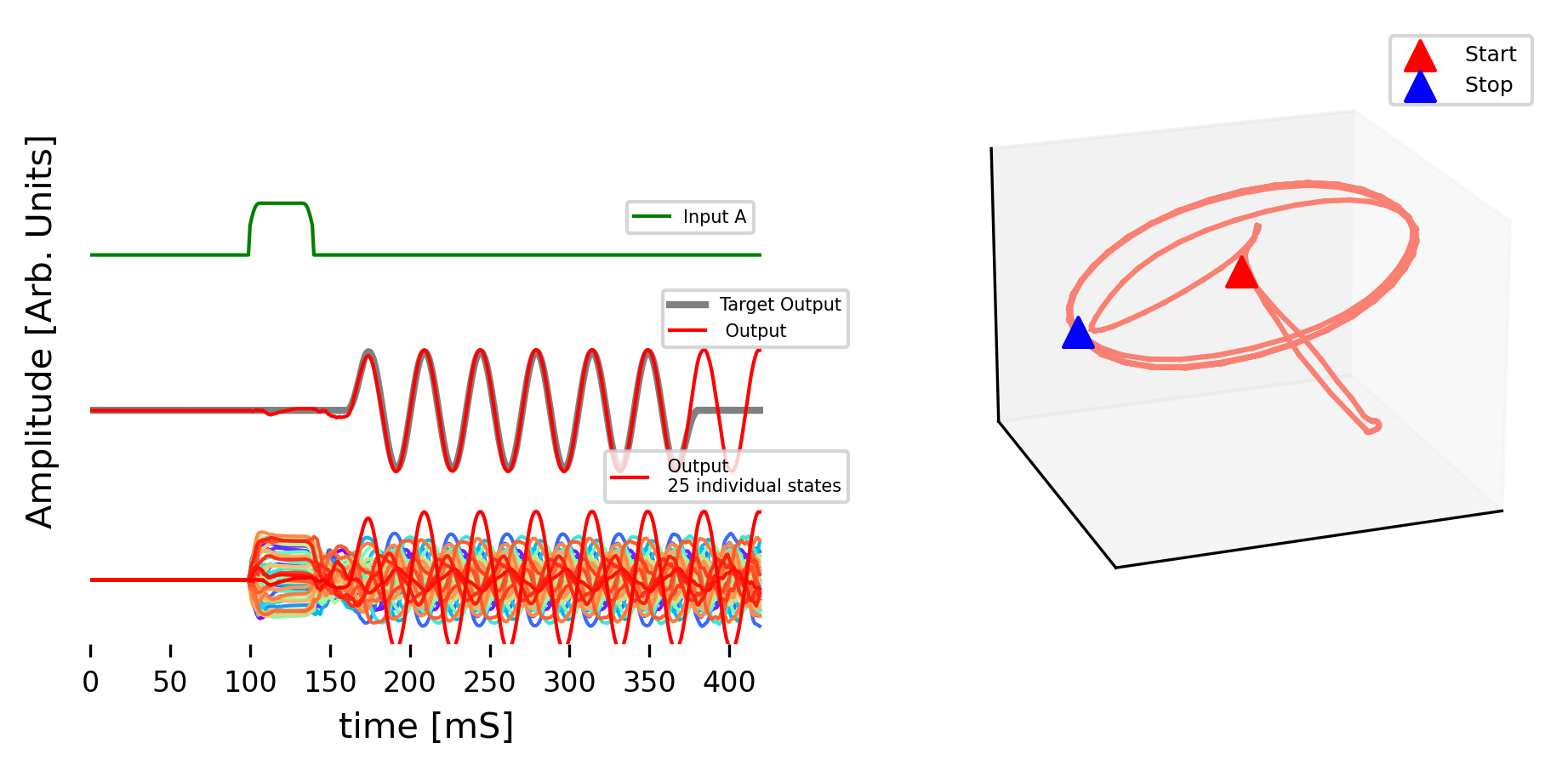}

\caption{{Neural Network Neural network $\#ID01$, trained for the Finite Duration oscilator task. Stimuli with and without noise applied on the same network produced a different effect.}}
\label{fig_10_c}
\end{center}
\end{figure}

{We believe this happens because the nature of this task is different from the decision-making tasks considered here. In this case, the network needs to learn to reproduce a pattern, and the presence or absence of input noise is a big difference for the network to generalize. This problem can be solved by including in the training data set samples that do not contain noise or including variations in the amplitude of the noise levels. But this will only make sense if the system to be parameterized and modelled presents such variation. Figure \ref{fig_10_c} shows an example of how the response of the same network varies when the stimulus of the network has noise (as it was trained) and when is stimulated with a signal without noise. In this second case, it is observed that the consequence is that the output continues to oscillate, which is not a behaviour for which it was trained.}

{These observations regarding what type of data sets to use have to do with how robust we want to design our systems and what properties we are seeking to represent and should be taken into account when comparing the abstract models studied here with those obtained from experimental data.}

Other examples of dynamics in the trained networks are available in the repository, for all studied tasks and different initial conditions, showing the different possible internal states achieved with different pre-training connectivity weights (Supplementary materials).

In this subsection, we showed that a small, fully connected nonlinear RNN trained with Adam and backpropagation through time can successfully learn and reproduce many different tasks. We also showed that the post-training phase space is not uniquely determined by the learned task, as different dynamical solutions (for the recurrent units) are compatible with a single learned behaviour (output unit), which is consistent with \cite{maheswaranathan2019universality}.

{In \cite{maheswaranathan2019universality} the authors found that the geometry of the representation of the RNN is highly sensitive to the choice of the different architectures (RNN, GRU or LSTM), which is expected since the equations of these architectures, and the internal states are different from each other. They found that while the geometry of the network can vary throughout the architectures, the topological structure of the fixed points, transitions between them, and the linearized dynamics appear universally in all architectures. However, in Maheswaranathan et al., it is not stated that each topological structure is linked uniquely to each task. The results of the present work do not contradict them. This analysis is an extension showing how, within the same network architecture (in this case RNNs), different topological structures can be obtained, as long as at least one structure (and one transition) is associated with the different decisions for which the network has been trained to take in the task.}

\subsection{Network memory capacity and size scaling} \label{pulse_studies}

\begin{figure}[!tbph]
\begin{center}
\includegraphics[totalheight=4.75cm]{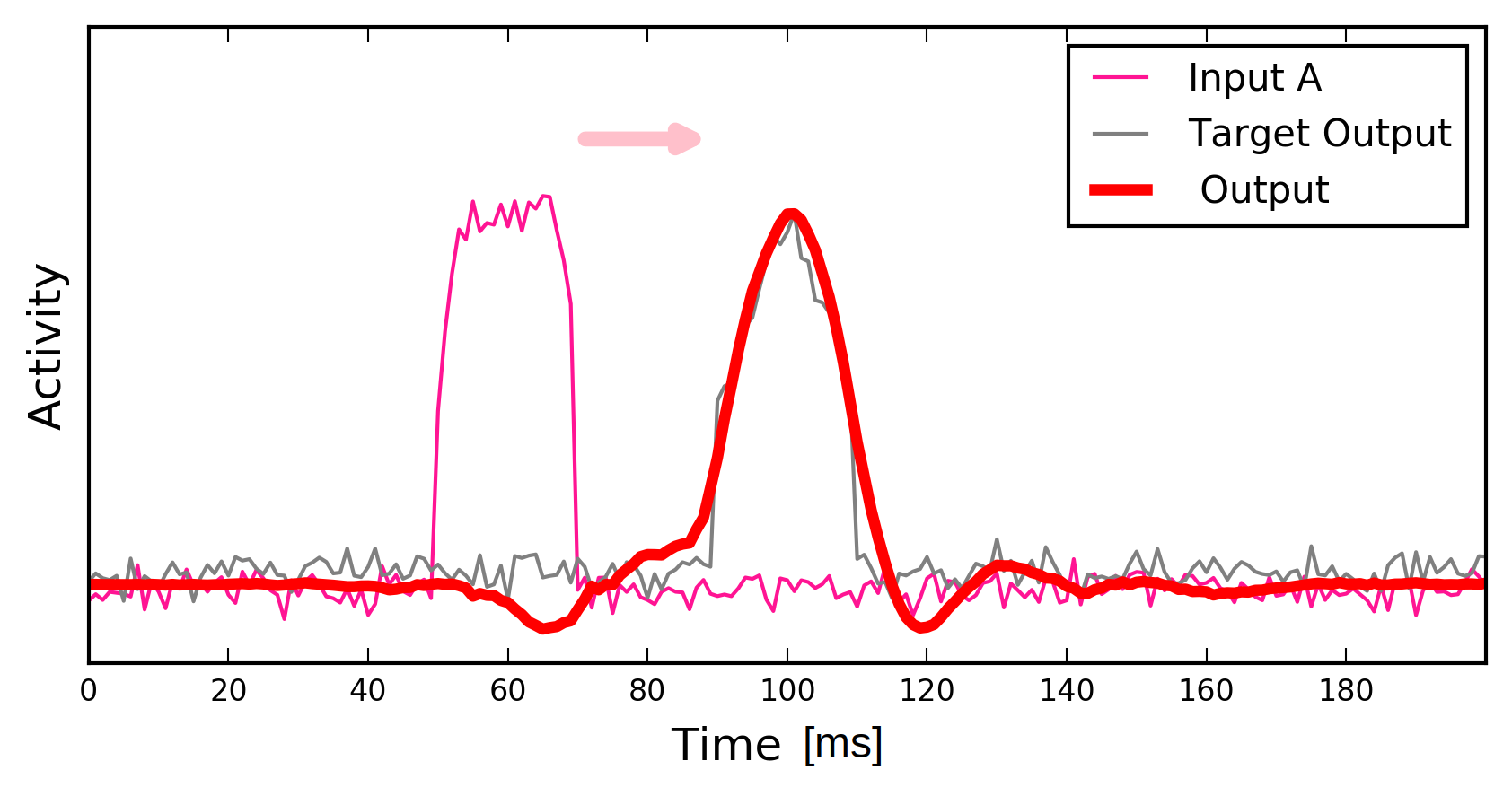}
\includegraphics[totalheight=4.75cm]{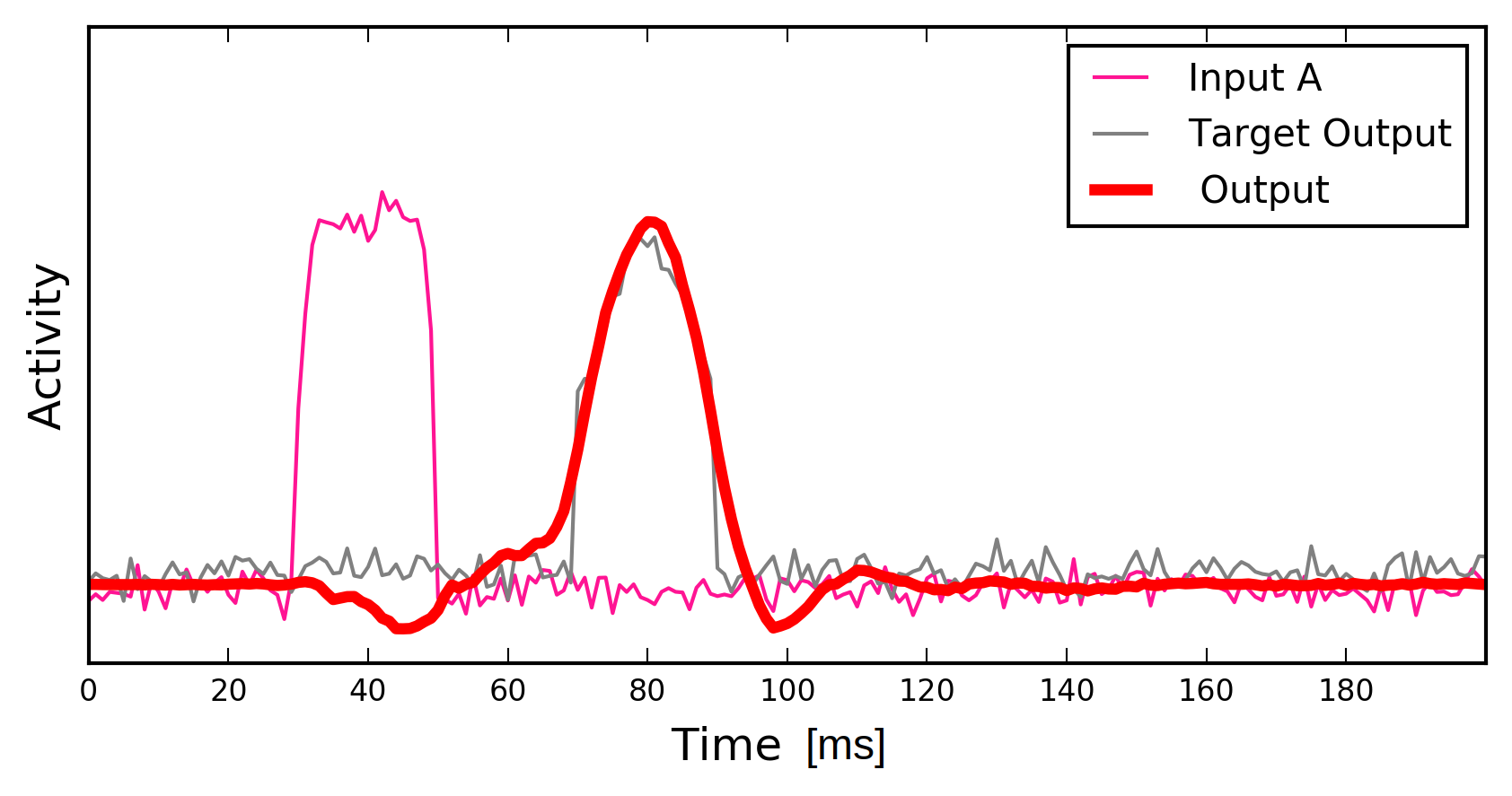}
\includegraphics[totalheight=4.75cm]{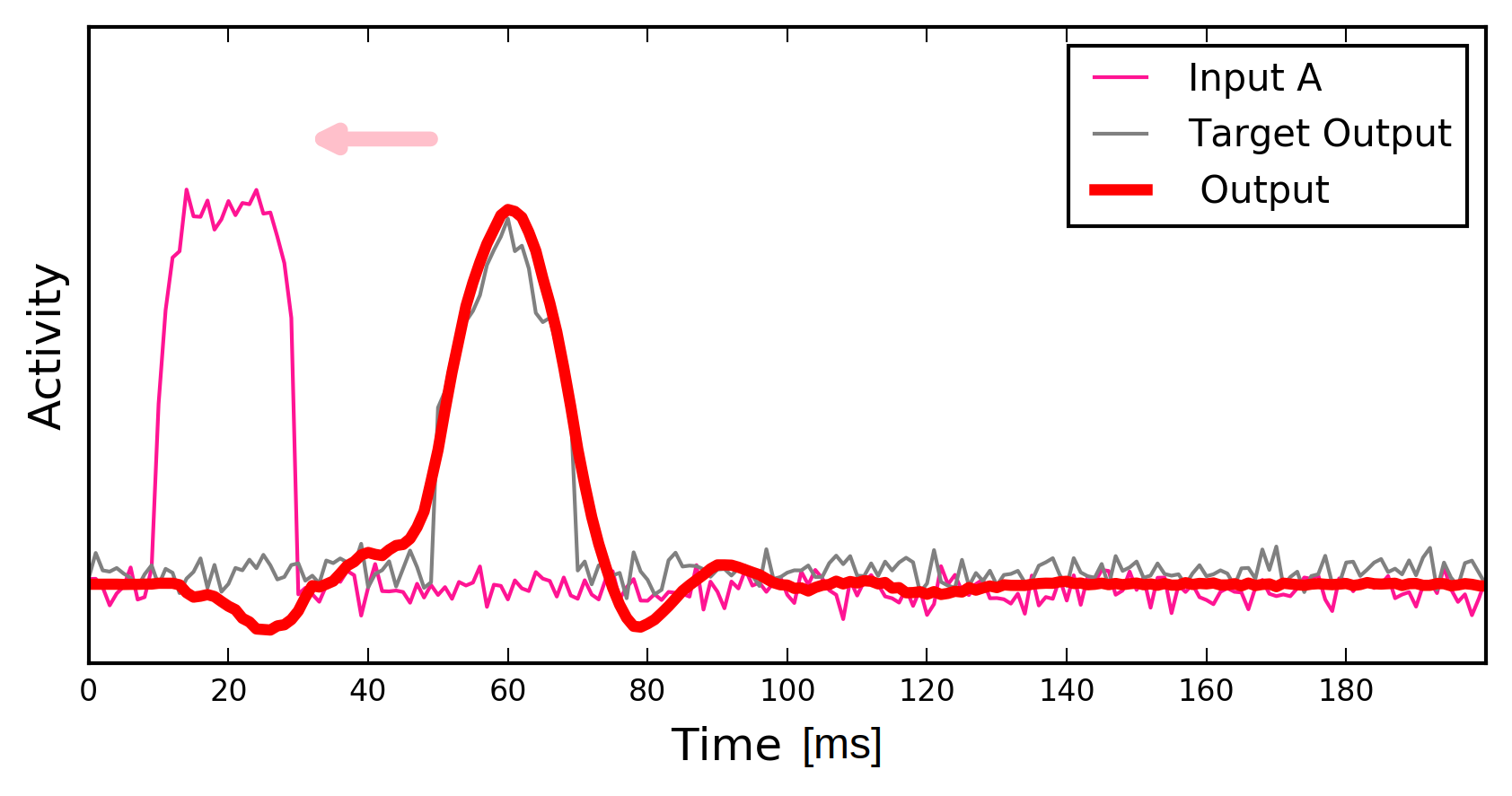}
\caption{Time transnational invariance for the stimulus for an ``Time reproduction" task. The trained network is stimulated with a time series where the stimulus pulse occurs in different moments. The pink line represents the state of the input signal. The Grey line represents the output response and the red thick line is the output target.}
\label{fig_13}
\end{center}
\end{figure}

Another interesting aspect of the trained networks is that they are translationally invariant in time, even though they were always trained with the stimulus occurring at the same moment. This situation is shown in Figure \ref{fig_13}. The network correctly responds to the input pulse no matter when it arrives. 

{This happens consistently for the AND, OR, and XOR decision-making tasks and for the oscillation generator. Invariance to translations in time does not happen for the NOT task. In this case, the task consists of learning not to react after a certain time, and the network cannot generalize to time intervals not seen for the chosen parameterization.} See the Supplementary Information for translational time invariance of the AND task. 

We asked ourselves how much time between the stimulus and the answer is possible to learn for a particular network size. This problem is related to the well-known vanishing gradient problem concerning long-time dependencies. When estimating the gradient for further minimization, by applying the chain rule the value of the gradient will show short- and long-time dependencies on past values. Long-time dependencies suffer from the shrinkage of the gradient (see Supplementary Material). In the context of ML, this problem was solved by using other recurrent network architectures such as LTSTM and GRU \cite{cho2014learning, Hochreiter:1997:LSM:1246443.1246450}.

{Vanishing gradient is not the only cause for the recurrent network failing to learn a task. By fixing the network`s size in terms of the number of units, we are fixing the number of parameters which are involved in gradient estimation. For a given number of epochs, and only changing the time delay, we are studying indirectly the effect in modeling the response to long-term dependencies.}

To show the temporal limitations of the model, we performed a study where we trained a set of networks on the time reproduction task, and then measure the rate of success in terms of Euclidean distance between target and output. {The distance between the network's predicted output and the target output was estimated using the Numpy function $linalg.norm()$, which in this case is the Frobenius norm (or euclidean norm) between the output vector of the trained network and the target output.}

\begin{figure}[htb!]
\begin{center}
\hspace*{-1cm}\includegraphics[width=11cm]{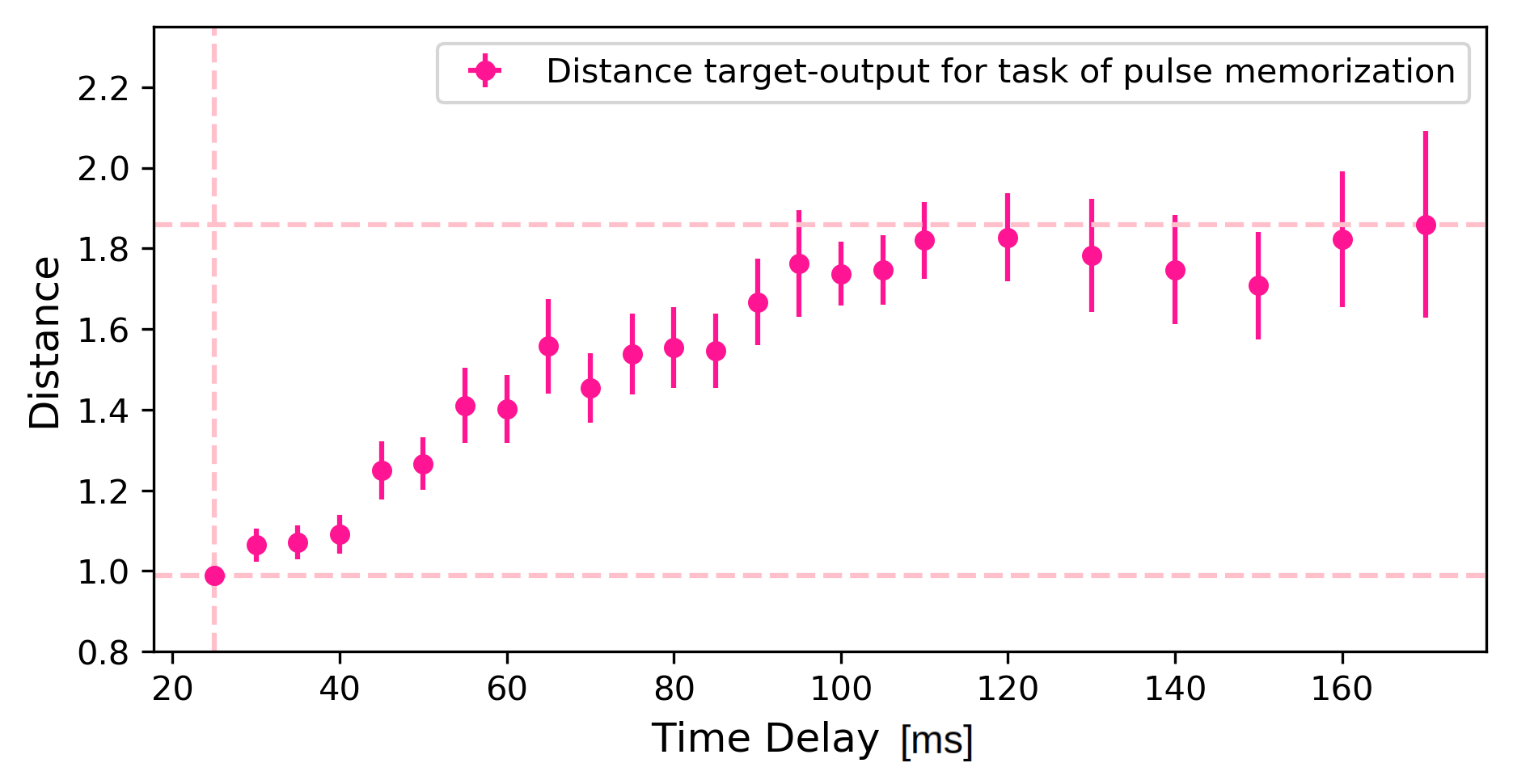}
\hspace*{-1cm}\includegraphics[width=11cm]{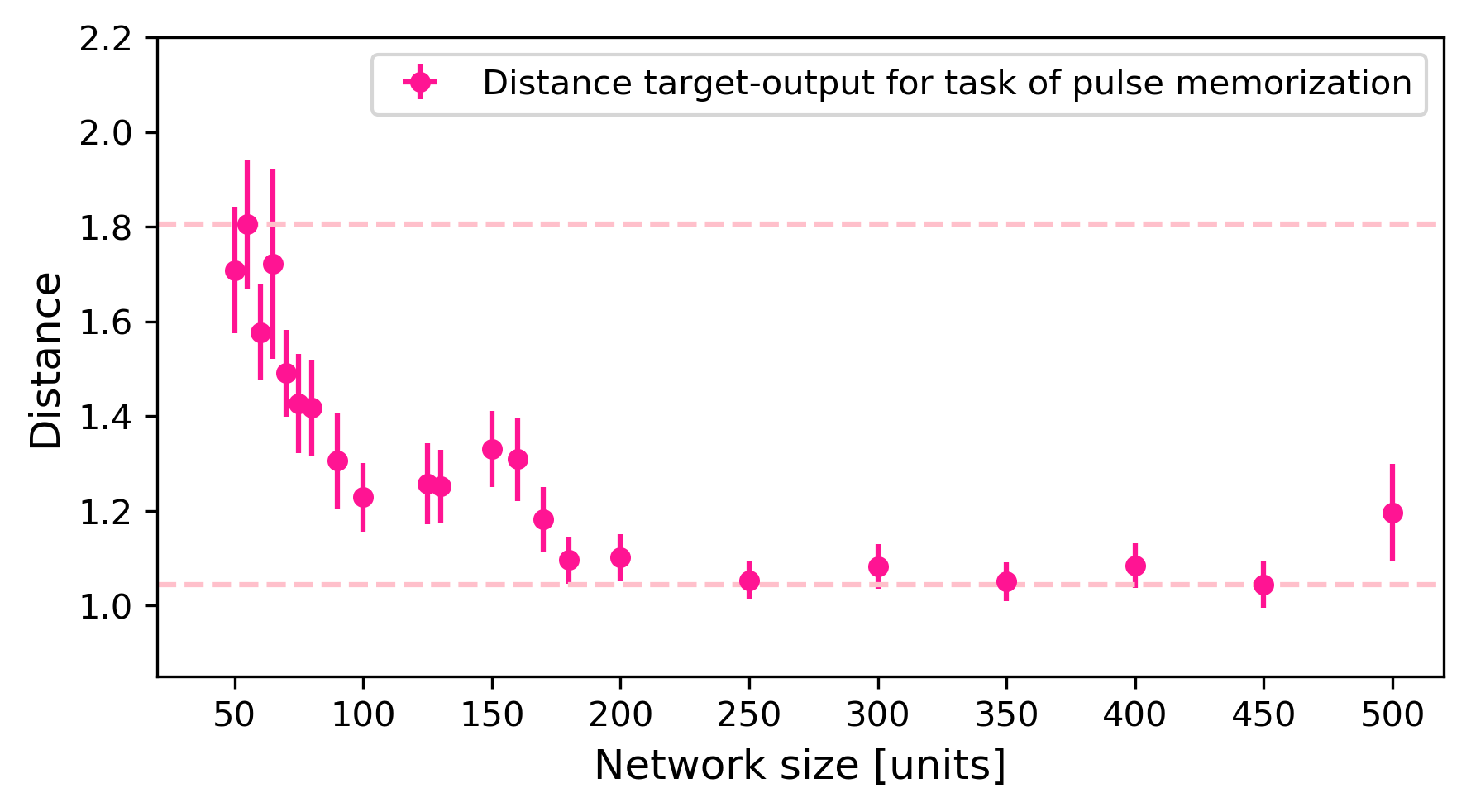}
\caption{{Target-output distance in the ``Time reproduction task” and scaling properties. Upper panel: Mean distance between stimulus and response of the learned task vs. time. Bottom panel: Mean distance between target and output vs. size of the network. A distance close to 1 means that all networks reproduce the desired output for the task. A distance close to 1.8, means that almost non of the networks could reproduce the task, given the worst-case distance between the target and output. Dashed pink lines are the maximum and minimum values for the distance.}}
\label{fig_14}
\end{center}
\end{figure}

The top panel of Figure \ref{fig_14} shows the result of our study. Each point in the plot is obtained as the average of the distance obtained for the considered set of 20 networks trained for that time interval of response. For this task and a particular duration of the time series, a distance close to 1 means that all networks reproduce the desired output for the task for the sample test set with good performance. When a distance is equal to 1.8, it means that almost none of the networks could reproduce the task given the worst-case distance between target and output, meaning the maximum possible distance between signal and target.

Our results presented in the upper panel of Figure \ref{fig_14} shows that the mean distance increases with the time duration until it reaches the worst performance at around 120 ms.

Next, we considered a fixed 150 ms delay between the input signal and response, where the success rate is low. We then increased the number of recurrent units. 

{When we increment the number of units, we are increasing the number of parameters of the model. In the context of Machine Learning has been proved that such a strategy improves learning, up to a certain point, depending also on the minimization algorithm \cite{DBLP:journals/corr/abs-1211-5063}}

The results are shown in the bottom panel of Figure \ref{fig_14}. For a fixed time interval, the memory capacity improves with the size of the network as expected, reaching the best performance when the number of units approaches 200. Larger networks do not give any additional advantage.

{Regarding increasing the number of epochs, it does not improve the training when the time delay is large. We performed the experiment changing this parameter for the "AND" with a long delay in time from 20 to 100, and there is no improvement in the network's performance of the trained networks.}

{Finally, regarding the amplitude of the pulses, it will depend on the training data set. The network is not able to generalize the response to any amplitude variation, but with our parameterization, networks are robust and can respond, for example, in the AND task to input pulses by performing the task when the amplitude of the input stimuli is reduced to $55\%$ of its initial value, or when it increases to $600\%$ (See Supplementary Information).}

\subsection{Network response to damage}\label{dam}

We induced post-training ``damage" to a network that was previously successfully trained by removing connections. We then measure the performance of the network as a function of the degree of damage.

{The damage done to a pre-trained neural network by removing connections is different from dropout \cite{10.5555/2627435.2670313} and pruning \cite{https://doi.org/10.48550/arxiv.1506.02158}. These techniques have other effects on the network's structure and performance have different aims and are performed during different moments. Dropout is applied during training.}

{Dropout means randomly dropping out neurons during training, which induces the network to learn more robust and generalizable features. Pruning, on the other hand, is also a regularization technique. It means removing the weakest connections in the network, which can lead to a more sparse and efficient network during training. Removing connections from a pre-trained network, which is the case here, has a different effect, diminishing the performance. These effects can be generalised to other decision-making tasks. We showed one task for simplicity but the effect is the same for the others. Here we are not studying training performance. We are quantifying how important the connection weights are according to sign and intensity between units in terms of the network being able to perform a task. We are trying to understand which connections are redundant or essential to the task performance in each state.}

{The damage studies on the trained network are aimed to disentangle whether the result of the training of the binary decision-making tasks could result in the clustering of the neurons in response to the different stimuli or not. In addition, we wanted to study whether there is a relationship between the strength of the damaged recurrent connections and the performance of a task. If there were clustering in the response of the neurons by removing certain units (removing all their connections and replacing with cero the values in the matrix), the network could have a different reaction to the four combinations of stimuli. It will respond with different accuracy in each case.}

{First, we tried the approach of randomly removing entire units to see if the network was still capable of performing the task or at least responding to any of the combinations of stimuli differently. But randomly removing even one unit  destroyed the learning in all the different responses to the combinations of input stimuli. This proves that the network responses are not clustered, in terms of neurons or connections. Next, we studied removing connectivities in terms of intensity, if there was a relationship between the response of the network to the different combinations of input stimuli and how strong or meaningful the connections could be.}

We considered a set of 10 successfully trained networks (50 recurrent units) that perform the AND task. Since we are using fully connected networks, the total number of connections is 2500, including positive and negative connections. 

We removed the connections gradually in a symmetric way by zeroing a growing number of the connections from the smallest (in absolute value) up in the connectivity distribution. For each percentage removed from each network, we calculated the distance between the target and the output in the four possible combinations of input values, and then the average. 

\begin{figure}[htb!]
\begin{center}
\textbf{a)}\includegraphics[width=5.78cm]{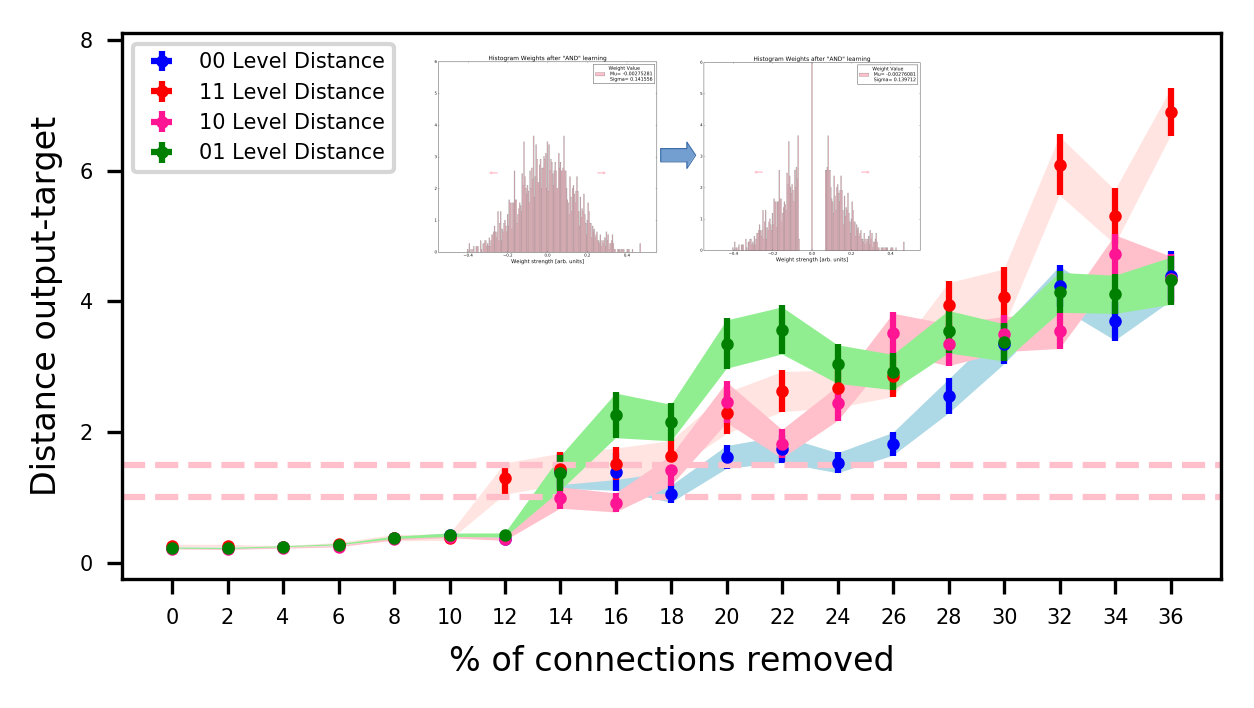}\hspace{-0.5cm}\textbf{b)}\includegraphics[width=5.78cm]{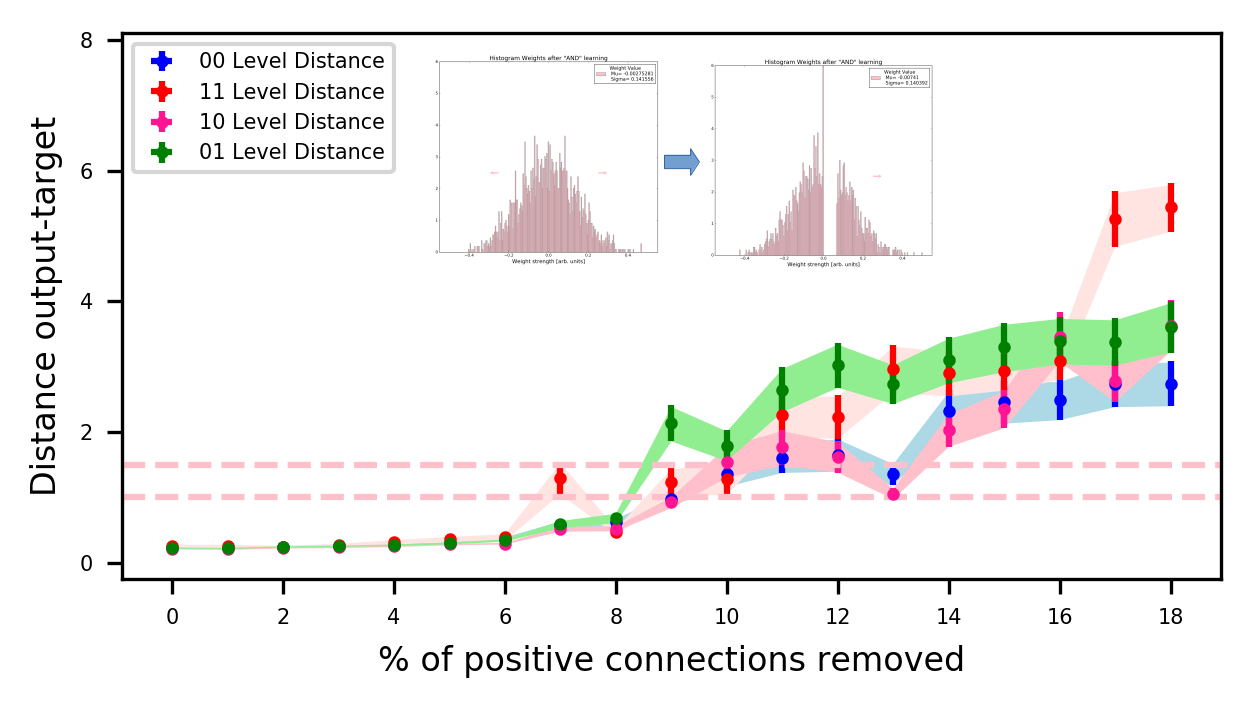} 

\textbf{c)}\includegraphics[width=5.78cm]{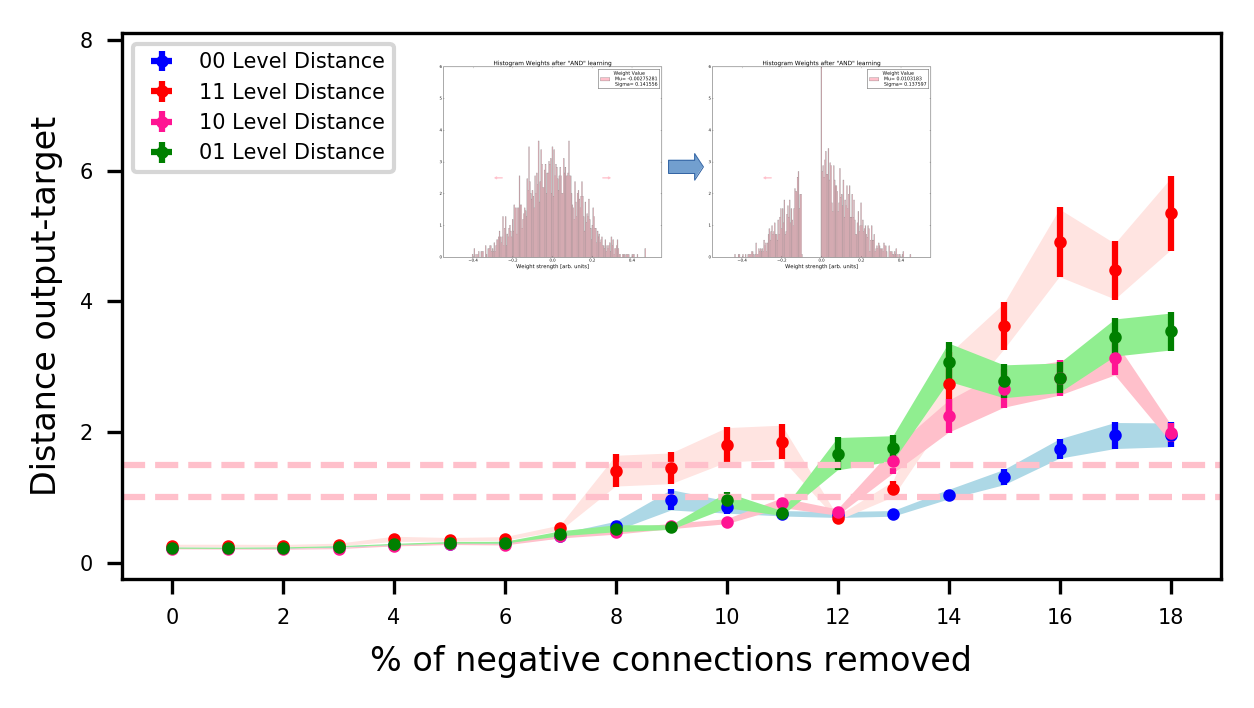} \hspace{-0.5cm}\textbf{d)}\includegraphics[width=5.78cm]{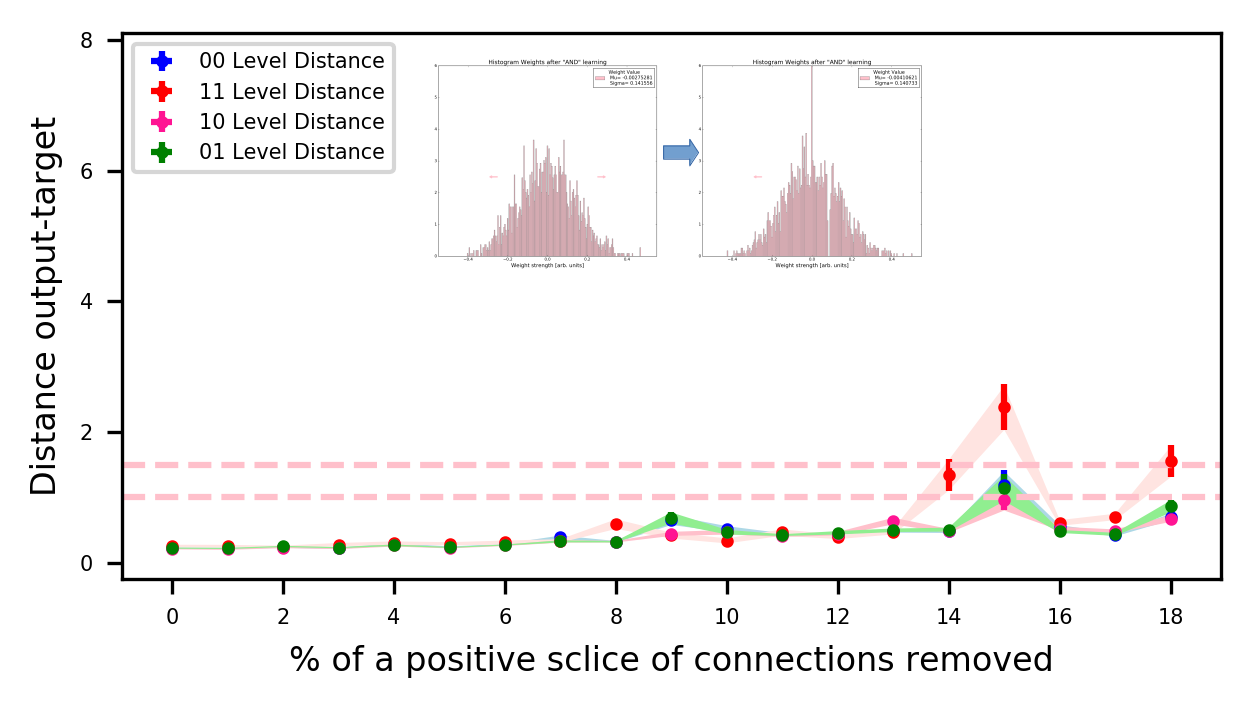} 

\textbf{e)}\includegraphics[width=5.78cm]{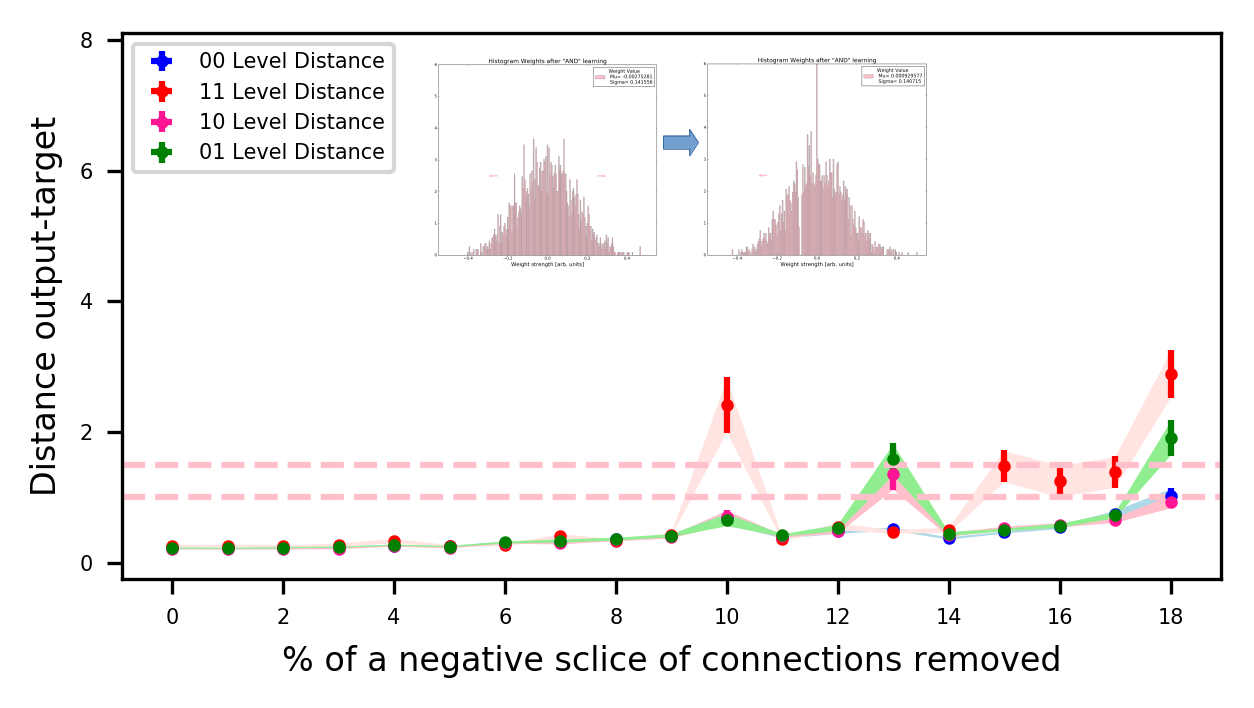}

\caption{Damage analysis: results of removing connections in RNN. \textbf{(a)} Removal of the smallest (in absolute value) connections up to the corresponding percentile. \textbf{(b)} and \textbf{(c)}) Same as (a), but only positive or negative connections, respectively. \textbf{(d)} and \textbf{(e)} Removal of connections within the corresponding percentile only (positive and negative connections, respectively). Mean $\pm$ standard error across 10 networks. Within each of the five figures is represented an example of how a band of weights is removed from the distribution of weight connections.}
\label{fig_16}
\end{center}
\end{figure}

The result of this study is shown in  Figure \ref{fig_16} \textbf{a}, trained and damaged connectivity distributions in the insets. Coloured lines represent the results for every input configuration of the AND task (Table \ref{tabla_and}). In this plot, all connections up to the given percentage are removed. It is clear that, when we removed all the connections (positive and negative) with strength in the lowest 14\%, the output of the networks deteriorates (the distance between output and target is larger than 1 for any input configuration). At 20\% of connections removed, the distances are larger than 1.5, meaning the networks stop working correctly (output very different from the target).

In panels \textbf{b} and \textbf{c} of Figure \ref{fig_16}, we show the result of removing either positive or negative only connections, respectively, up to the given percentage. Here we show that the networks are disrupted, if we remove either only positive or only negative connections. Both types are equally necessary to perform the considered task, and there is no apparent difference for either sign, which is consistent with our generic networks (no distinct excitatory and inhibitory subpopulations). 

Next we study what happens if we remove  a single percentile (Figure \ref{fig_16} \textbf{d} and \textbf{e}; for instance, 14\% means that all connections between 13\% and 14\% are zeroed). The output deteriorates at values a little higher than before. These results show that both the connectivity strength and the number of removed connections are important.

The effect of removing connections on the network causes the learned task to deteriorate or destroy, whether one removes bands of connections of a certain intensity above a value, or if one accumulates the removal of many connections. The effect does not depend on whether we remove negative or positive connections. The cumulative effect of removing the lower connection portion produces deterioration at a lower percentage intensity value.

{Finally, we summarise the results of this study. With the training method used in the present work, without applying regularization mechanisms and considering simple decision-making tasks, the responses to stimuli are not clustered nor do they depend on the sign of connectivity. Also, removing bands of connectivity values of intensity is almost as detrimental as removing connectivities below a certain threshold.}

\section{Discussion}\label{discu}

Decision-making in time processing tasks involves the perception, production, comparison, and maintenance of time intervals in working memory \cite{Bi10530}. These processes are crucial for animals to anticipate or act correctly at the right time. Neural network models are useful to understand computation in this context.

Training recurrent networks to imitate different bio-inspired or cognitive tasks is not new.
However, in the literature, there are not many examples of open-source frameworks to use {until recently}. Some of them are {\cite{Yang2019ER, molano22}}. Two examples of useful code that can be used also are: \cite{DBLP:journals/neco/SussilloB13, 10.1371/journal.pcbi.1004792}. In \cite{DBLP:journals/neco/SussilloB13} the main focus is the dynamics and fixed pints, and in \cite{10.1371/journal.pcbi.1004792} RNNs with a particular structure are considered.

{Our work also is a new example of an open-source framework that can be used and modified to include different aspects of task parametrization. 
Particularly, it shows how the parameterization of the tasks is crucial in the different emergent properties in the trained networks.}

Also, within the decision-making tasks, a subset of those that we proposed had not been implemented and characterized before. We obtained different realizations for the same task with different possible dynamical behaviours, which is consistent with \cite{maheswaranathan2019universality} as indicated in Section \ref{dyn}.

{We have previously studied how single tasks can be combined via a contextual signal and briefly explored some properties of multitasking \cite{Jarne_2021}. In this work, we have studied in more detail the activity of the networks and the consequences of the different ways of parameterizing and modelling decision-making or pattern-generation tasks.}

{The brain encodes sensory stimuli through the collective activity of thousands of neurons. The coding process in this high-dimensional space is typically studied using linear decoding and dimensionality-reduction techniques such as those presented here. The underlying network is often described as a dynamical system \cite{BARAK20171, SUSSILLO2014156, 10.1371/journal.pcbi.1009271}. The decision-making options from tasks are reflected in the eigenvalue spectrum with the values outside the unit circle.}

{Our results in this work suggest that additional information is necessary to constrain the models and be able to compare them thoroughly. For example, different topological structures can be obtained within the same initial network architecture as long as at least one structure (set of dominant eigenvalues) is associated with the different decisions or final states for which the network has been trained. We have to be able to analyze which solutions are motivated by results from biology.}

{Regularization methods could be used to penalize some solution types against others if one had some argument or hypothesis related to the dynamics of the biological process that motivates it. Because we implemented a task-based approach, in principle, for the trained networks, nothing restricts the different possible internal configurations that give rise to the same decision-making process.}

{One explanation for the emergence of the multiplicity of solutions in trained networks is related to the simplicity of the tasks considered and the absence of regularization mechanisms imposed during training. For the same set of initial conditions, the training can converge to different solutions that are equivalent in readout activity but different in recurrent activity.}

To our knowledge, this is the first time where initialization differences in trained RNNs for bio-inspired tasks are studied by comparing the Orthogonal Initial condition and Random Normal initial condition.

{A very particular initialization scheme with excellent performance is to initialize a recurrent neural network with an identity matrix, i.e., a matrix with ones on the diagonal and zeros elsewhere \cite{https://doi.org/10.48550/arxiv.1504.00941}, which translates in all eigenvalues initially equal to one. This can be useful in the field of machine learning. One of the primary reasons is that it can help alleviate the vanishing or exploding gradient problem, which is a common issue in training deep neural networks. The identity matrix allows the gradients to propagate through time without significantly amplifying or attenuating them, thereby improving the stability and speed of training. Additionally, the use of an identity matrix as the initial weight matrix can help ensure that the network starts with a balanced and symmetric configuration, which can help improve its overall performance and ability to learn complex patterns. Overall, initializing a recurrent neural network with an identity matrix can be a useful technique for improving its stability, performance, and ability to learn. But, from the point of view of Computational Neuroscience, such configurations are arbitrary and far from the biological characteristics desired to endow the models.}

This is the first detailed study on the network scale in terms of temporal response and size concerning the capacity to learn such tasks.

Also, this is the first study performed on the trained RNN with damage to provide insights into how robust is a trained network in terms of its connectivity. A better understanding of model constraints for simple tasks, such as the ones studied in the present work, could help to develop better models in computational neuroscience.

{Different temporal tasks could require more than encoding time and can have distinct computational requirements, which include exhibiting temporal scaling, generalising to novel contexts, or robustness to noise. This work helps to understand how RNNs can encode time response and satisfy distinct computational requirements, but we also knowledge that neural activity at the population level can exhibit different computational or generalization properties to consider.}

{Understanding which features arise as a result of choices for task parameterization and which properties are related to behaviour in cortical areas from data is a challenge. This work aims to help this by characterising how robust the responses in the activity are when the input stimuli vary, when different response times are considered, when parameters or  hyperparameters of the network (such as size) are changed, or when the characteristics of the trained network are perturbed. This work provides, on the one hand, constraints on the models, but on the other, gives ideas on how to define more adequately the task parameterization under study so that it is adapted to the characteristics of the tasks in the laboratory with which individuals or animals are trained.}

\section{Conclusions} \label{conclu}

We have presented the results of a set of studies performed on RNNs trained to perform various temporal and flow control tasks. We showed that small-sized networks with simple rate models for the individual units are adequate to learn and perform tasks in response to temporally dynamic inputs. We also showed that recurrent networks can learn a given task by developing different internal dynamics---for instance, a constant value in the output can be produced by the recurrent network either converging to a fixed point or entering a limit cycle \cite{doi:10.1146/annurev-neuro-092619-094115}. 

With this study, we were able to characterize the memory limits for a given trained network, showing a trade-off between network size and target duration for a simple task. We explicitly showed how the problem of the vanishing gradients arises as the target duration is increased, which would help when selecting a specific model, network size, and target time scale.

We showed how much damage can sustain a trained network before collapsing. In our model, it is the cumulative effect of removing connections that have a greater effect, rather than the value of the largest connections removed. In other words, we observed, given this training scheme and topology, in which way the task is broken when deactivating certain parts of the network and which part of the weights is significant. We also showed that the responses to stimuli are not clustered nor do they depend on the sign of connectivity.

{The three proposed analyses are valuable when constructing neural network models employed in Computational Neuroscience. In this work, we analyzed a simple model of a small network, making available our framework for its use in further neuroscience studies whenever task parameterization needs to be designed.}

{One must be cautious when interpreting model results from RNNs simulations due to the multiplicity of possible outcomes. To provide an example of a possible wrong conclusion that can be drawn from brain modelling, one might assume that the observed network behaviour accurately reflects the underlying biology of the modelled brain region. }{ However, our work shows that network training methods can lead to a variety of responses that do not necessarily could reflect typical cortical phenomena. Therefore, our contribution is to highlight the importance of considering appropriate data-driven hypotheses when developing a truly descriptive model. In this context, we are referring to hypotheses about the behaviour and dynamics of the modelled brain region, such as whether it exhibits oscillatory or non-oscillatory idle states.}

We showed that certain characteristics can emerge naturally when applying network training methods and that the responses obtained due to them are varied and do not necessarily reflect phenomena typical of the cortex, but rather typical of the model used. 

{Further steps in our studies will include constraints motivated by the biology of the brain, such as a distinction between excitatory and inhibitory units and the study of other cognitive tasks similar to Context-dependent decision-making making and different kinds of working memory tasks.}

\section{Supplementary Information}\label{sup-a}

Github repository: \\

\url{https://github.com/katejarne/RNN\_study\_with\_keras}\\

{Suplementary text in pdf.}

\section*{Ethical Approval}

Not applicable.
 
\section*{Competing interests}

The authors declare that they have no known competing financial interests or personal relationships that could have appeared to influence the work reported in this paper.
 
\section*{Authors' contributions}

C.J. designed the original version of the research, developed the code, performed simulations, analyzed data and wrote the manuscript. R.L. supervised research, suggested PC analysis of neural trajectories, network size study and damage study visualization, and edited the manuscript.
 
\section*{Funding}

Work is supported by CONICET and UNQ. C. Jarne acknowledge support from PICT 2020-01413.
 
\section*{Availability of data and materials}

Code, simulations, and additional figures of this analysis are available at the following Github repository: \\

\url{https://github.com/katejarne/RNN\_study\_with\_keras}\\

\section*{Acknowledgments}

Present work was supported by CONICET and UNQ. C. Jarne acknowledge support from PICT 2020-01413. We want to thank also the anonymous reviewers for their careful reading of the manuscript and their insightful comments and suggestions.





\bibliography{sn-bibliography.bib}


\end{document}